\documentclass[12pt,english,manuscript,aps,prd,A4,english,12pt]{revtex4-1}
\usepackage[T1]{fontenc}
\usepackage[latin9]{inputenc}
\usepackage{color}
\usepackage{babel}
\usepackage{amsmath}
\usepackage{amssymb}
\usepackage{graphicx}
\usepackage{esint}
\usepackage[unicode=true,pdfusetitle,
 bookmarks=true,bookmarksnumbered=false,bookmarksopen=false,
 breaklinks=false,pdfborder={0 0 1},backref=false,colorlinks=true]
 {hyperref}
\hypersetup{
 pdfborderstyle=,linkcolor=blue,citecolor=cyan}

\makeatletter

\providecommand{\tabularnewline}{\\}

\usepackage{babel}

\usepackage{xcolor}

\makeatother

\begin{document}
\title{Analytic solutions of relativistic dissipative spin hydrodynamics
\\
 with radial expansion in Gubser flow}
\author{Dong-Lin Wang}
\email{wdl252420@mail.ustc.edu.cn}

\affiliation{Department of Modern Physics, University of Science and Technology
of China, Anhui 230026, China}
\author{Xin-Qing Xie}
\email{xq936400957@mail.ustc.edu.cn}

\affiliation{Department of Modern Physics, University of Science and Technology
of China, Anhui 230026, China}
\author{Shuo Fang}
\email{fangshuo@mail.ustc.edu.cn}

\affiliation{Department of Modern Physics, University of Science and Technology
of China, Anhui 230026, China}
\author{Shi Pu}
\email{shipu@ustc.edu.cn}

\affiliation{Department of Modern Physics, University of Science and Technology
of China, Anhui 230026, China}
\begin{abstract}
We have derived the analytic solutions of dissipative relativistic
spin hydrodynamics with Gubser expansion. Following the standard strategy
of deriving the solutions in a Gubser flow, we take the Weyl rescaling
and obtain the energy-momentum and angular momentum conversation equations
in the $dS_{3}\times\mathbb{R}$ space-time. We then derive the analytic
solutions of spin density, spin potential and other thermodynamic
in $dS_{3}\times\mathbb{R}$ space-time and transform them back into
Minkowski space-time $\mathbb{R}^{3,1}$. In the Minkowski space-time,
the spin density and spin potential including the information of radial
expansion decay as $\sim L^{-2}\tau^{-1}$ and $\sim L^{-2}\tau^{-1/3}$
in large $L$ limit, with $\tau$ being proper time and $L$ being
the characteristic length of the system, respectively. Moreover, we
observe the non-vanishing spin corrections to the energy density and
other dissipative terms in the Belinfante form of dissipative spin
hydrodynamics. Our results can also be used as test beds for future
simulations of relativistic dissipative spin hydrodynamics. 
\end{abstract}
\maketitle

\section{Introduction}

\label{sec:introduction}


A large amount of orbital angular momentum perpendicular to the reaction
plane is generated in non-central relativistic heavy-ion collisions
at Relativistic Heavy Ion Collider (RHIC). Due to the spin-orbit coupling,
particles produced in collisions are polarized along the direction
of initial angular momentum. Such kinds of polarization can be measured
through the weak decay of $\Lambda$ and $\overline{\Lambda}$ hyperons
\citep{ZTL_XNW_2005PRL,ZTL_XNW_2005PLB,Gao:2007bc}. The global polarization
of $\Lambda$ and $\overline{\Lambda}$ hyperons measured by STAR
Collaboration \citep{STAR:2007ccu,STAR:2017ckg,STAR:2018gyt} confirms
the early theoretical predictions \citep{ZTL_XNW_2005PRL,ZTL_XNW_2005PLB,Gao:2007bc}
and is described by phenomenological models well \citep{Becattini:2007nd,Becattini:2007sr,Becattini:2013fla,Becattini:2013vja,Fang_2016PRC_polar,Karpenko_epjc2017_kb,XieYilong_prc2017_xwc,LiHui_prc2017_lpwx,Sun:2017xhx,Shi_plb2019_sll,WeiDexian_prc2019_wdh,Fu:2020oxj,Ivanov:2020wak,Ryu:2021lnx,Lei:2021mvp}.
One can also see the recent reviews \citep{Becattini:2020ngo,Gao:2020vbh,Liu:2020ymh}
and references therein. Very recently, the global spin polarization
at low energy region attracts lots of attention and needs to be systematically
studied \citep{Ivanov_prc2019_its,Deng:2020ygd,STAR:2021beb,Guo:2021udq,Ayala:2021xrn,Deng:2021miw}.


Furthermore, the STAR experiments have also measured the local spin
polarization of $\Lambda$ and $\overline{\Lambda}$ hyperons along
the beam and out-of-plane directions \citep{Niida:2018hfw,Adam:2019srw}.
The experimental data shows that the local spin polarization along
the beam direction as a function of the azimuthal angle is opposite
to the theoretical predictions from both transport and hydrodynamic
models \citep{Karpenko_epjc2017_kb,XieYilong_prc2017_xwc,Becattini_prl2018_bk,Florkowski_prc2019_fkmr,Wu:2019eyi,Fu:2020oxj}.
Such difference can not be explained by feed-down effects \citep{Xia:2019fjf,Becattini_epjc2019_bcs,Li:2021jvn}.
It is referred as the ``sign problem'' in the local spin polarization.
There are also many other approaches, e.g. by considering hadronic
interactions \citep{Csernai:2018yok,Nogueira-Santos:2020aky}, quantum
kinetic approaches \citep{Gao:2019znl,Weickgenannt:2019dks,Hattori:2019ahi,Wang:2019moi,Sheng:2020oqs,Weickgenannt:2020aaf,Guo:2020zpa,Yang:2020hri,Liu:2020flb,Manuel:2021oah,Sheng:2021kfc,Weickgenannt:2021cuo,Lin:2021mvw},
simulation for the chiral kinetic theory with side-jump effects \citep{Liu:2019krs}
and relativistic spin hydrodynamics \citep{Florkowski:2017ruc,Florkowski:2018fap,Montenegro:2018bcf,Hattori:2019lfp,Bhadury:2020puc,Bhadury:2020cop,Garbiso:2020puw,Gallegos:2020otk,Shi:2020htn,Fukushima:2020ucl,Li:2020eon,Bhadury:2021oat,Gallegos:2021bzp,She:2021lhe,Peng:2021ago,Hongo:2021ona,Wang:2021ngp}.
Although some approaches \citep{Liu:2019krs,Voloshin_2018epjWeb}
agree with the data qualitatively and certain progress has been made,
this problem has not been solved completely till now and requires
further investigation.

One possible problem related to the sign problem is that the spin
degrees of freedom may not reach the global equilibrium so that the
Cooper-Frye formula \citep{Becattini:2013fla,Fang_2016PRC_polar}
at global equilibrium fails to reproduce local spin polarization.
It suggests that we need to add the spin degree of freedom to the
current phenomenological frameworks and consider the off-equilibrium
effects \citep{Adam:2019srw,Becattini:2020ngo,Liu:2020ymh,Gao:2020vbh,Liu:2021uhn,Liu:2021nyg,Fu:2021pok,Becattini:2021suc,Becattini:2021iol}.



Very recently, the shear-induced polarization (SIP) as one of off-equilibrium
effects \citep{Liu:2021uhn,Liu:2021nyg,Fu:2021pok,Becattini:2021suc,Becattini:2021iol}
has been proposed and plays an important role to the spin polarization.
The numerical results from relativistic hydrodynamics including SIP
can give a correct sign in the comparison with the experimental data
in the named the strange quark equilibrium scenarios \citep{Fu:2021pok},
or in the isothermal equilibrium scenarios \citep{Becattini:2021iol}.
On the other hand, the total polarization in strange quark equilibrium
scenarios is also found to be sensitive to the equation of state,
freeze-out temperature and other parameters \citep{Yi:2021ryh}. Similar
studies on the parameter dependence at the $\sqrt{s_{NN}}=19.6\textrm{GeV}$
collisions are shown in Ref. \citep{Sun:2021nsg} and one can also
see Ref. \citep{Florkowski:2021xvy,Ryu:2021lnx,Alzhrani:2022dpi, Yi:2021unq,Wu:2022mkr}
for other related discussions. It indicates that the off-equilibrium
effects need to be systematically studied in the future. Moreover,
the modified Cooper-Frye formula including off-equilibrium effects,
such as the effects of collisions \citep{Shi:2020qrx} and spin \citep{Liu:2021nyg}
has been discussed.


On the other hand, there are two general ways to add the spin degree
of freedom to the current phenomenological frames. As a macroscopic
effective theory, the relativistic spin hydrodynamics is one possible
way to consider the spin effects in the heavy ion collisions. In ordinary
relativistic hydrodynamics, spin is not encoded into the conservation
equations of energy, momentum and net charge or baryon number. To
describe the influence of spin, there are several ways to construct
the relativistic spin hydrodynamics, such as through effective Lagrangian
approaches \citep{Montenegro:2017rbu,Montenegro:2017lvf}, entropy
current analysis \citep{Hattori:2019lfp,Fukushima:2020ucl,Li:2020eon,She:2021lhe,Gallegos:2021bzp,Hongo:2021ona},
quantum statistical operators approach \citep{Becattini:2007nd,Becattini:2009wh,Becattini_plb2019_bfs,Becattini:2020sww},
and quantum kinetic theory \citep{Florkowski:2017ruc,Florkowski:2018fap,Florkowski:2018myy,Hattori:2019ahi,Bhadury:2020puc,Shi:2020qrx,Peng:2021ago}.
Meanwhile, the microscopic descriptions for spin dynamics is the quantum
kinetic theories for massive fermions with collisions \citep{Gao:2019znl,Weickgenannt:2019dks,Weickgenannt:2020aaf,Weickgenannt:2020sit,Hattori:2019ahi,Yang:2020hri,Liu:2020flb,Weickgenannt:2021cuo,Sheng:2021kfc,Fang:2022ttm},
which is a natural extension of chiral kinetic theory \citep{Stephanov:2012ki,Son:2012zy,Chen:2012ca,Manuel:2013zaa,Manuel:2014dza,Chen:2014cla,Chen:2015gta,Hidaka:2016yjf,Mueller:2017lzw,Hidaka:2018mel,Huang:2018wdl,Gao:2018wmr,Liu:2018xip,Lin:2019ytz,Lin:2019fqo}.
Also, see recent reviews \citep{Wang:2017jpl,Becattini:2020ngo,Gao:2020vbh,Bhadury:2021oat,Hidaka:2022dmn}
and the reference therein.


Moreover, in recent studies \citep{Shi:2020htn,Buzzegoli:2021wlg,Liu:2021nyg},
a modified Cooper-Frye formula with spin potential at local equilibrium
has been derived, in which the spin potential, just like the thermal
vorticity and shear viscous tensor, contributes to spin polarization
pseudo-vector.

Although there are intensive discussions on relativistic spin hydrodynamics,
the codes for the numerical simulations have not been developed yet.
It results in the lack of decaying behavior of spin density and spin
potential. To see the influence of spin potential in the modified
Cooper-Frye formula \citep{Buzzegoli:2021wlg,Liu:2021nyg}, we need
to know the decaying behavior of these terms. Meanwhile, the numerical
simulations also require some analytic solutions in special configurations
as the test-beds.


To estimate the decay behavior of spin potential and find the suitable
test-beds for the future numerical simulations, we search for the
analytic solutions of relativistic spin hydrodynamics at some certain
configurations. Based on the canonical form of relativistic spin hydrodynamics
\citep{Hattori:2019lfp,Fukushima:2020ucl,Hongo:2021ona}, we have
already derived the analytic solutions in Bjorken expansion \citep{Wang:2021ngp}.
Our results show that the spin density and spin potential decay as
$\tau^{-1}$ and $\tau^{-1/3}$, respectively, where $\tau$ is the
proper time. We also find that only one component of spin density,
$S^{xy}$, do not accelerate the Bjorken velocity. The transverse
expansion of the medium and other components of spin density are not
allowed in our previous study \citep{Wang:2021ngp}.


In this work, we search for the analytical solutions of relativistic
dissipative spin hydrodynamics in Gubser expansion by following the
similar strategy for the relativistic magnetohydrodynamics \citep{Pu:2016ayh,Roy:2015kma,Pu:2016bxy,Pu:2016rdq,Siddique:2019gqh,Wang:2020qpx}.
The Gubser flow \citep{Gubser:2010ze,Gubser:2010ui,Pang:2014ipa,Nopoush:2014qba,Yan:2015lfa,Martinez:2017pvs,Martinez:2017ibh,Behtash:2017wqg,Chattopadhyay:2018pwe,Calzetta:2019dfr,Behtash:2019qtk,Dash:2020zqx,Singh:2020rht},
which can describe the radial expansion, is more closer to the reality
in heavy-ion collisions. As expected, our analytical solution in a
Gubser flow will contain the information of transverse expansion.

We emphasize that we derive the analytic solutions of relativistic
spin hydrodynamics with radial expansion in a Gubser flow and it is
different with some other approaches, in which the Bjorken or Gubser
expansion is treated as the expanding background \citep{Florkowski:2019qdp,Singh:2020rht,Singh:2021man,Florkowski:2021wvk}.


The paper is organized as follows: In Sec. \ref{sec:Gubser-Flow},
we review the basic idea of Gubser flow. In Sec. \ref{sec:Relativistic-dissipative-spin},
we introduce the canonical form of spin hydrodynamics, the conservation
equations and constitutive relations in both Minkowski space $\mathbb{R}^{3,1}$
and $dS_{3}\times\mathbb{R}$ space-time. In Sec. \ref{sec:Analytic-solution-in},
we simplify the differential equations in a Gubser flow and derive
the analytic solutions. We also discuss the results in the Belinfante
form of spin hydrodynamics in Sec. \ref{sec:Results-for-Belinfante}.
We concludes and summarizes this work in Sec. \ref{sec:Conclusion}.

Throughout this work, we choose the metric $g_{\mu\nu}$ in Minkowski
space-time with Cartesian coordinates as $\mathrm{diag}\{-,+,+,+\}$,
the normalized fluid velocity $u^{\mu}$ satisfying $u_{\mu}u^{\mu}=-1$
and the transverse projection operator $\Delta^{\mu\nu}\equiv g^{\mu\nu}+u^{\mu}u^{\nu}$.
In curved space-time, we often use the covariant derivative $\nabla_{\mu}$
while $\partial_{\mu}$ denotes the ordinary derivative. In addition,
the symbols $\partial_{\perp}^{\mu}\equiv\Delta^{\mu\nu}\partial_{\nu}$
and $\nabla_{\perp}^{\mu}\equiv\Delta^{\mu\nu}\nabla_{\nu}$ represent
projection derivatives. For simplicity, we define the symmetric and
antisymmetric parts of an arbitrary tensor $A^{\mu\nu}$ as $A^{(\mu\nu)}\equiv(A^{\mu\nu}+A^{\nu\mu})/2$
and $A^{[\mu\nu]}\equiv(A^{\mu\nu}-A^{\nu\mu})/2$ . When considering
the viscous tensor, we also introduce the symmetric and traceless
part of $A^{\mu\nu}$, $A^{<\mu\nu>}\equiv\frac{1}{2}\left[\Delta^{\mu\alpha}\Delta^{\nu\beta}+\Delta^{\nu\alpha}\Delta^{\mu\beta}\right]A_{\alpha\beta}-\frac{1}{3}(\Delta^{\alpha\beta}A_{\alpha\beta})\Delta^{\mu\nu}$.
To avoid the misleading, a physical variable $A$ with a hat, i.e.
$\hat{A}$, denotes that it is defined in $dS_{3}\times\mathbb{R}$
space-time.

\section{Review on Gubser flow \label{sec:Gubser-Flow}}

In this section, following Refs. \citep{Gubser:2010ze,Gubser:2010ui}
we briefly review the main results in a Gubser flow. Besides the Bjorken
boost invariance, Gubser flow can describe expansion with azimuthal
symmetry in transverse plane \citep{Gubser:2010ze,Gubser:2010ui}.

Following \citep{Gubser:2010ze,Gubser:2010ui}, one can construct
Gubser flow by imposing the ``Gubser symmetry'' $SO(3)\times SO(1,1)\times Z_{2}$,
which strongly restricts the profile of fluid velocity. It is challenging
to find the fluid velocity satisfying Gubser symmetry in Minkowski
space-time $\mathbb{R}^{3,1}$. Fortunately, one can solve the hydrodynamic
equations in the manifold $dS_{3}\times\mathbb{R}$, in which $dS_{3}$
refers to the 3-dimensional de Sitter space-time, and transform the
solution back to Minkowski space-time $\mathbb{R}^{3,1}$ through
Weyl rescaling.

We rewrite the metric in Minkowski space-time $\mathbb{R}^{3,1}$
with coordinates $(t,x,y,z)$ as 
\begin{eqnarray}
ds^{2} & = & -dt^{2}+dx^{2}+dy^{2}+dz^{2}\nonumber \\
 & = & -d\tau^{2}+dx_{\perp}^{2}+x_{\perp}^{2}d\varphi^{2}+\tau^{2}d\eta^{2},\label{eq: ds in R31}
\end{eqnarray}
where 
\begin{eqnarray}
x=x_{\perp}\cos\varphi, & \qquad & y=x_{\perp}\sin\varphi,\nonumber \\
t=\tau\cosh\eta, & \qquad & z=\tau\sinh\eta.\label{eq: CT-1}
\end{eqnarray}
The coordinates $\tau,x_{\perp},\varphi$ and $\eta$ denote longitudinal
proper time, transverse plane radius, azimuthal angle and rapidity,
respectively. We then introduce a time-like hyperbola embedding into
the manifold $\mathbb{R}^{3,1}$. The radius of this hyperbola is
normalized to be $1$, i.e. 
\begin{eqnarray}
-X_{0}^{2}+X_{1}^{2}+X_{2}^{2}+X_{3}^{2} & = & 1,
\end{eqnarray}
where $X_{\mu}$ is the Cartesian coordinates in $\mathbb{R}^{3,1}$and
can be parametrized as: 
\begin{eqnarray}
X_{0}=-\frac{L^{2}-\tau^{2}+x_{\perp}^{2}}{2\tau L}, & \qquad & X_{1}=\frac{x_{\perp}\cos\varphi}{\tau},\nonumber \\
X_{3}=\frac{L^{2}+\tau^{2}-x_{\perp}^{2}}{2\tau L}, & \qquad & X_{2}=\frac{x_{\perp}\sin\varphi}{\tau}.\label{eq:CartesianCoordinatesInR31}
\end{eqnarray}
Here, $L$ is an adjustable parameter with dimension of length. The
line element of $dS_{3}$ can be expressed as, 
\begin{eqnarray}
ds_{3}^{2} & = & \frac{1}{\tau^{2}}\left(-d\tau^{2}+dx_{\perp}^{2}+x_{\perp}^{2}d\varphi^{2}\right).
\end{eqnarray}
Now, the metric of Minkowski space-time $\mathbb{R}^{3,1}$ can be
transformed into the one of $dS_{3}\times\mathbb{R}$ under Weyl rescaling
with factor $\tau$, 
\begin{eqnarray}
d\hat{s}^{2}\equiv\frac{1}{\tau^{2}}ds^{2} & = & \frac{1}{\tau^{2}}\left(-d\tau^{2}+dx_{\perp}^{2}+x_{\perp}^{2}d\varphi^{2}\right)+d\eta^{2},\label{eq:metric_ds3_R_01}
\end{eqnarray}
where $ds^{2}$ is given by Eq. (\ref{eq: ds in R31}). Notice that
Weyl rescaling is not a coordinate transformation \citep{Gubser:2010ui}.
Later, the metric (\ref{eq:metric_ds3_R_01}) is rewritten by the
Gubser coordinates $(\rho,\theta,\varphi,\eta)$, 
\begin{eqnarray}
d\hat{s}^{2} & = & -d\rho^{2}+\cosh^{2}\rho\left(d\theta^{2}+\sin^{2}\theta d\varphi^{2}\right)+d\eta^{2},
\end{eqnarray}
where 
\begin{eqnarray}
\sinh\rho & = & -\frac{L^{2}-\tau^{2}+x_{\perp}^{2}}{2L\tau},\nonumber \\
\tan\theta & = & \frac{2Lx_{\perp}}{L^{2}+\tau^{2}-x_{\perp}^{2}},\label{eq: CT-2}
\end{eqnarray}
Calculations in current work are performed mainly in this coordinates
unless otherwise specified.

Now, we discuss the velocity profile in a Gubser flow. The Gubser
symmetry requires that the normalized velocity must be 
\begin{equation}
\hat{u}_{\mu}=\left(-1,0,0,0\right),\label{eq:fluid_velocity_01}
\end{equation}
in $dS_{3}\times\mathbb{R}$ \citep{Gubser:2010ze,Gubser:2010ui},
\cite{Aaboud:2018eph}which means the Gubser flow is static in the
$(\rho,\theta,\varphi,\eta)$ coordinate system. Under the Weyl rescaling
$u_{\mu}=\tau\hat{u}_{\mu}$ and coordinate transformation, one can
derive the fluid velocity in the $(\tau,x_{\perp},\varphi,\eta)$
coordinates in Minkowski space-time $\mathbb{R}^{3,1}$, 
\begin{eqnarray}
u_{\mu} & = & \left(-\frac{1}{\cosh\rho}\frac{L^{2}+\tau^{2}+x_{\perp}^{2}}{2L\tau},\frac{1}{\cosh\rho}\frac{x_{\perp}}{L},0,0\right),\label{eq:velocity in Minkowski}
\end{eqnarray}
where $\rho$ is given by Eq. (\ref{eq: CT-2}). Different with the
standard Bjorken flow, velocity in a Guber flow depends on the $x_{\perp}$.

Meanwhile, it can be proved that the fluid velocity (\ref{eq:fluid_velocity_01})
holds during the space-time evolution of the Gubser flows, i.e. the
space-time covariant derivatives of fluid velocity (\ref{eq:fluid_velocity_01})
always vanish in a Gubser flow.

\section{Relativistic dissipative spin hydrodynamics\label{sec:Relativistic-dissipative-spin}}

In this section, we extend the main equations in the relativistic
dissipative spin hydrodynamics from Minkowski space-time $\mathbb{R}^{3,1}$
\citep{Hattori:2019lfp,Fukushima:2020ucl,Li:2020eon,She:2021lhe,Hongo:2021ona}
to $dS_{3}\times\mathbb{R}$ space-time. Note that, the energy momentum
conservation and spin evolution equations are not invariant under
Weyl rescaling \citep{Singh:2020rht,Singh:2021liw}.

\subsection{Relativistic dissipative spin hydrodynamics in Minkowski space $\mathbb{R}^{3,1}$\label{subsec:Relativistic-dissipative-spin}}

The canonical form of energy momentum tensor $T^{\mu\nu}$ for dissipative
spin hydrodynamics in Minkowski space-time reads \citep{Hattori:2019lfp,Fukushima:2020ucl,Hongo:2021ona},
\begin{eqnarray}
T^{\mu\nu} & = & \left(e+p\right)u^{\mu}u^{\nu}+pg^{\mu\nu}+2h^{\left(\mu\right.}u^{\left.\nu\right)}+\text{\ensuremath{\pi^{\mu\nu}}}+T^{[\mu\nu]},\label{eq:EMT_Min}
\end{eqnarray}
with energy density $e$, pressure $p$, heat flux $h^{\mu}$ and
viscosity tensor $\text{\ensuremath{\pi^{\mu\nu}}}$. The antisymmetric
part $T^{[\mu\nu]}$ is further decomposed as, 
\begin{equation}
T^{[\mu\nu]}=2q^{\left[\mu\right.}u^{\left.\nu\right]}+\phi^{\mu\nu},\label{eq:anti_EMT_Min}
\end{equation}
where $q^{\mu}\equiv u_{\alpha}\Delta_{\beta}^{\mu}T^{[\alpha\beta]}$
and $\phi^{\mu\nu}\equiv\Delta_{\alpha}^{\mu}\Delta_{\beta}^{\nu}T^{[\alpha\beta]}$.
The $q^{\mu}$ and $\phi^{\mu\nu}$ play a role of the source to produce
or absorb the spin.

The main equations for dissipative spin hydrodynamics are 
\begin{eqnarray}
\nabla_{\mu}T^{\mu\nu} & = & 0,\label{eq:energy-momentum conservation}\\
\nabla_{\alpha}\Sigma^{\alpha\mu\nu} & = & -2T^{[\mu\nu]},\label{eq:spin evolution}
\end{eqnarray}
where $\Sigma^{\alpha\mu\nu}$ is the rank-three canonical spin components
in the total angular momentum. Note that we have replaced the ordinary
derivative $\partial_{\mu}$ with the covariant derivative $\nabla_{\mu}$
for the general space-time. The Eq.(\ref{eq:spin evolution}) comes
from the total angular momentum conservation and describes the spin
evolution. Furthermore, one can decompose $\Sigma^{\alpha\mu\nu}$
as \citep{Hattori:2019lfp,Fukushima:2020ucl,Hongo:2021ona}, 
\begin{eqnarray}
\Sigma^{\alpha\mu\nu} & = & u^{\alpha}S^{\mu\nu}+\Sigma_{(1)}^{\alpha\mu\nu},\label{eq:sigma_01}
\end{eqnarray}
where $S^{\mu\nu}=-S^{\nu\mu}$ is named as the spin density and $\Sigma_{(1)}^{\alpha\mu\nu}$
is perpendicular to the fluid velocity $u_{\alpha}\Sigma_{(1)}^{\alpha\mu\nu}=0$.
The above decomposition for $\Sigma^{\alpha\mu\nu}$ is called the
a non-anti-symmetric gauge which has been used in Refs. \citep{Montenegro:2017rbu,Florkowski:2017ruc,Hattori:2019lfp,Fukushima:2020ucl,Li:2020eon,Gallegos:2021bzp,She:2021lhe}
and also in spin hydrodynamics for massless fermions \citep{Fukushima:2018osn}.
One can construct a total anti-symmetric tensor for the spin density,
which has been used in Refs. \citep{Bhadury:2020puc,Hongo:2021ona}.

We regard $S^{\mu\nu}$ as an independent variable and introduce the
corresponding spin potential $\omega^{\mu\nu}=-\omega^{\nu\mu}$,
which is conjugate to spin density $S^{\mu\nu}$. To add the effects
of spin, the thermodynamic relations become, 
\begin{eqnarray}
e+p & = & Ts+\omega_{\mu\nu}S^{\mu\nu},\nonumber \\
dp & = & sdT+S^{\mu\nu}d\omega_{\mu\nu},\label{eq: thermodynamic relations in R31}
\end{eqnarray}
where $T$ and $s$ denote the temperature and entropy density, respectively.
For simplicity, here we set the particle number density and the charge
or baryon chemical potential are always zero throughout the current
work. To highlight the spin effect, we also neglect the heat flux
$h^{\mu}$ in the energy-momentum tensor $T^{\mu\nu}$ in Eq. (\ref{eq:EMT_Min}).

The power counting scheme can be assumed as $\omega_{\mu\nu}\sim\mathcal{O}\left(\partial^{1}\right)$,
$S^{\mu\nu}\sim\mathcal{O}\left(\partial^{0}\right)$ and $\Sigma_{(1)}^{\alpha\mu\nu}\sim\mathcal{O}\left(\partial^{1}\right)$
\citep{Fukushima:2020ucl} or $\omega_{\mu\nu}\sim\mathcal{O}\left(\partial^{1}\right)$,
$S^{\mu\nu}\sim\mathcal{O}\left(\partial^{1}\right)$ and $\Sigma_{(1)}^{\alpha\mu\nu}\sim\mathcal{O}\left(\partial^{2}\right)$
\citep{Hattori:2019lfp,Hongo:2021ona}.

The spin evolution equation (\ref{eq:spin evolution}) becomes, 
\begin{eqnarray}
\nabla_{\alpha}\left(u^{\alpha}S^{\mu\nu}\right) & = & -4q^{\left[\mu\right.}u^{\left.\nu\right]}-2\phi^{\mu\nu}.\label{eq:spin evolution-1}
\end{eqnarray}
According to the second law of thermodynamics \citep{Hattori:2019lfp,Fukushima:2020ucl}
or the effective theories \citep{Hongo:2021ona}, constitutive equations
are given by, 
\begin{eqnarray}
\pi^{\mu\nu} & = & -\eta_{s}\nabla{}^{\left\langle \mu\right.}u^{\left.\nu\right\rangle }-\zeta\Delta^{\mu\nu}\nabla_{\alpha}u^{\alpha},\nonumber \\
q^{\mu} & = & -\lambda\left(\frac{1}{T}\nabla_{\perp}^{\mu}T-u^{\alpha}\nabla_{\alpha}u^{\mu}-4\omega^{\mu\nu}u_{\nu}\right),\nonumber \\
\phi^{\mu\nu} & = & -\gamma\left(\nabla_{\perp}^{\left[\mu\right.}u^{\left.\nu\right]}-2\Delta^{\mu\alpha}\Delta^{\nu\beta}\omega_{\alpha\beta}\right).\label{eq:constitutive relation}
\end{eqnarray}
where $\eta_{s}$ and $\zeta$ are the shear viscosity and bulk viscosity,
respectively and $\lambda,\gamma$ are two new transport coefficients
related to the spin. The four coefficients $\eta_{s},\zeta,\lambda,\gamma$
in the constitutive relations are all non-negative.

\subsection{Main equations in $dS_{3}\times\mathbb{R}$ space-time \label{subsec:Main-equations-in}}

In this subsection, we transform the spin hydrodynamic equations in
the previous subsection to the $dS_{3}\times\mathbb{R}$ space-time.

The Weyl rescaling in Eq. (\ref{eq:metric_ds3_R_01}) implies that
$g_{\mu\nu}\rightarrow\hat{g}_{\mu\nu}=\frac{1}{\tau^{2}}g_{\mu\nu}$
and $g^{\mu\nu}\rightarrow\hat{g}^{\mu\nu}=\tau^{2}g^{\mu\nu}$. The
Christoffel symbols $\hat{\Gamma}_{\mu\nu}^{\lambda}$ in $dS_{3}\times\mathbb{R}$
space-time are related to the one in the Minkowski space-time $\Gamma_{\mu\nu}^{\lambda}$
by the following relation \citep{Wald:1984rg,Carroll:2004st,Singh:2020rht}
\begin{eqnarray}
\Gamma_{\mu\nu}^{\lambda} & = & \hat{\Gamma}_{\mu\nu}^{\lambda}+\frac{1}{\tau}\left(\delta_{\nu}^{\lambda}\hat{\nabla}_{\mu}\tau+\delta_{\mu}^{\lambda}\hat{\nabla}_{\nu}\tau-\hat{g}_{\mu\nu}\hat{g}^{\lambda\alpha}\hat{\nabla}_{\alpha}\tau\right),\label{eq:christoffel transform}
\end{eqnarray}
where $\hat{\nabla}_{\mu}$ is also defined in the $dS_{3}\times\mathbb{R}$
space-time. Similarly, we introduce the energy-momentum tensor in
the $dS_{3}\times\mathbb{R}$ space-time, 
\begin{equation}
\hat{T}^{\mu\nu}\equiv\tau^{\alpha}T^{\mu\nu},
\end{equation}
with $\alpha$ being a constant. The energy-momentum conservation
(\ref{eq:energy-momentum conservation}) becomes, \citep{Wald:1984rg}:
\begin{eqnarray}
\nabla_{\mu}T^{\mu\nu} & = & \partial_{\mu}T^{\mu\nu}+\Gamma_{\mu\lambda}^{\mu}T^{\lambda\nu}+\Gamma_{\mu\lambda}^{\nu}T^{\mu\lambda}\nonumber \\
 & = & \tau^{-\text{\ensuremath{\alpha}}}\left[\hat{\nabla}_{\mu}\hat{T}^{\mu\nu}-2\tau^{-1}\hat{T}^{[\mu\nu]}\hat{\nabla}_{\mu}\tau\right]
 +\tau^{-\text{\ensuremath{\alpha}}-1}\left[\left(6-\alpha\right)\hat{T}^{\lambda\nu}\hat{\nabla}_{\lambda}\tau-\hat{T}_{\;\mu}^{\mu}\hat{g}^{\nu\alpha}\hat{\nabla}_{\alpha}\tau\right].
\end{eqnarray}
Obviously, only when $\alpha=6$ and $\hat{T}_{\;\mu}^{\mu}=\tau^{\alpha}T_{\;\mu}^{\mu}=0$
is traceless, the last term proportional to $\tau^{-\alpha-1}$ vanishes.
The traceless condition of $\hat{T}^{\mu\nu}$ is satisfied in a conformal
fluid \citep{Wess:1971eb,Jarvis:2005hp,Baier:2007ix,Bhattacharyya:2007vs,Bhattacharyya:2008mz,Erdmenger:2008rm,Marrochio:2013wla},
in which the bulk viscosity is zero $\zeta=0$ and the $e=3p$ .

For simplicity, following the common strategy in a Gubser flow \citep{Gubser:2010ze,Gubser:2010ui},
we choose 
\begin{equation}
\alpha=6,
\end{equation}
and set $\zeta=0$ from now on. The energy-momentum conservation (\ref{eq:energy-momentum conservation})
then reduces to, 
\begin{eqnarray}
\hat{\nabla}_{\mu}\hat{T}^{\mu\nu}-2\tau^{-1}\hat{T}^{[\mu\nu]}\hat{\nabla}_{\mu}\tau & = & 0.\label{eq:EM conservation in ds3}
\end{eqnarray}

In general, the transformation rule for a physical variable $\hat{A}$
in $dS_{3}\textbackslash times\mathbb{R}$ space-time to its corresponding
quantity $A$ in Minkowski space-time is \citep{Baier:2007ix,Bhattacharyya:2007vs,Loganayagam:2008is,Gubser:2010ui},
\begin{eqnarray}
\hat{A}_{\nu_{1}...\nu_{n}}^{\mu_{1}...\mu_{m}}(x) & = & \tau^{\Delta_{A}}A_{\nu_{1}...\nu_{n}}^{\mu_{1}...\mu_{m}}(x),\label{eq:transformation rules}
\end{eqnarray}
where $\Delta_{A}=[A]+m-n$, and $[A]$ is the mass dimension of $A$.

Next, we discuss the spin evolution equation (\ref{eq:spin evolution-1}).
From Eq. (\ref{eq:christoffel transform}), the spin evolution equation
(\ref{eq:spin evolution-1}) is not covariant under Weyl rescaling.
In $dS_{3}\textbackslash times\mathbb{R}$ space-time, Eq. (\ref{eq:spin evolution-1})
becomes, 
\begin{eqnarray}
\hat{\nabla}_{\alpha}\left(\hat{u}^{\alpha}\hat{S}^{\mu\nu}\right) & = & \left(\hat{u}_{\alpha}\hat{S}^{\alpha\nu}\hat{g}^{\mu\beta}+\hat{u}_{\alpha}\hat{S}^{\mu\alpha}\hat{g}^{\nu\beta}+\hat{u}^{\mu}\hat{S}^{\nu\beta}-\hat{u}^{\nu}\hat{S}^{\mu\beta}\right)\tau^{-1}\hat{\nabla}_{\beta}\tau\nonumber \\
 &  & -4\hat{q}^{\left[\mu\right.}\hat{u}^{\left.\nu\right]}-2\hat{\phi}^{\mu\nu}.\label{eq:spin evolution in ds3}
\end{eqnarray}
where we have used that 
\begin{eqnarray}
\nabla_{\alpha}\left(u^{\alpha}S^{\mu\nu}\right) & = & \tau^{-6}\hat{\nabla}_{\alpha}\left(\hat{u}^{\alpha}\hat{S}^{\mu\nu}\right)-\tau^{-7}\left(\hat{u}_{\alpha}\hat{S}^{\alpha\nu}\hat{g}^{\mu\beta}+\hat{u}_{\alpha}\hat{S}^{\mu\alpha}\hat{g}^{\nu\beta}\right)\hat{\nabla}_{\beta}\tau\nonumber \\
 &  & -\tau^{-7}\left(\hat{u}^{\mu}\hat{S}^{\nu\beta}-\hat{u}^{\nu}\hat{S}^{\mu\beta}\right)\hat{\nabla}_{\beta}\tau.
\end{eqnarray}
Eq. (\ref{eq:spin evolution in ds3}) has several new terms proportional
to $\tau^{-1}\hat{\nabla}_{\beta}\tau$.

In this subsection, we extend the energy-momentum and angular momentum
conservation equations to the $dS_{3}\times\mathbb{R}$ space-time.
Unfortunately, these two kinds of conversation equations are not conformal
invariant, i.e. we find that there are extra terms $\sim\hat{\nabla}_{\mu}\tau$
in Eqs. (\ref{eq:EM conservation in ds3}, \ref{eq:spin evolution in ds3})
under Weyl rescaling.

Here, we emphasize that we do NOT consider the conformal fluid in
current study. In an ordinary fluid without spin, the anti-symmetric
part of energy-momentum tensor $T^{[\mu\nu]}$ is zero and Eq. (\ref{eq:EM conservation in ds3})
reduces to the simplest expression $\hat{\nabla}_{\mu}\hat{T}^{\mu\nu}=0$.
However, as discussed in Sec. \ref{subsec:Relativistic-dissipative-spin},
the anti-symmetric part of energy-momentum tensor $T^{[\mu\nu]}$
is non-vanishing in the spin hydrodynamics. We keep the general expression
(\ref{eq:EM conservation in ds3}) for energy-momentum conservation
here. Later, we will solve the conservation equations (\ref{eq:EM conservation in ds3})
in Sec. \ref{sec:Analytic-solution-in}.

\subsection{Constitutive equations in $dS_{3}\times\mathbb{R}$ space-time \label{subsec:Constitutive-relations-in}}

In this subsection, we discuss the constitutive equations (\ref{eq:EMT_Min},
\ref{eq:constitutive relation}) in $dS_{3}\times\mathbb{R}$ space-time.
The decomposition of $\hat{T}^{\mu\nu}$ in $dS_{3}\times\mathbb{R}$
space-time is similar to Eq. (\ref{eq:EMT_Min}, \ref{eq:anti_EMT_Min}),
\begin{eqnarray}
\hat{T}^{\mu\nu} & = & \left(\hat{e}+\hat{p}\right)\hat{u}^{\mu}\hat{u}^{\nu}+\hat{p}\hat{g}^{\mu\nu}+2\hat{h}^{\left(\mu\right.}\hat{u}^{\left.\nu\right)}+\text{\ensuremath{\hat{\pi}^{\mu\nu}}}+2\hat{q}^{\left[\mu\right.}\hat{u}^{\left.\nu\right]}+\hat{\phi}^{\mu\nu}.\label{eq: T^=00003D00005Cmu=00003D00005Cnu-in ds3}
\end{eqnarray}
By using Eq. (\ref{eq:transformation rules}), we find that $\hat{u}^{\mu}=u^{\mu}/\tau$,
$\hat{\pi}^{\mu\nu}=\tau^{6}\pi^{\mu\nu}$ and thermodynamic variables
become 
\begin{equation}
\hat{e}=\tau^{4}e,\qquad\hat{T}=\tau T,\qquad\hat{s}=\tau^{3}s.\label{eq:energy_density_ds3}
\end{equation}

Applying the Eq. (\ref{eq:christoffel transform}), the transformation
of $\nabla_{\mu}u^{\nu}$ is given by \citep{Bhattacharyya:2007vs}
\begin{eqnarray}
\nabla_{\mu}u^{\nu} & = & \tau^{-1}\hat{\nabla}_{\mu}\hat{u}^{\nu}+\tau^{-2}\delta_{\mu}^{\nu}\hat{u}^{\lambda}\hat{\nabla}_{\lambda}\tau-\tau^{-2}\hat{u}_{\mu}\hat{g}^{\nu\alpha}\hat{\nabla}_{\alpha}\tau.\label{eq:transformation of u}
\end{eqnarray}
It is straightforward to show that the bulk viscosity term $\zeta\Delta^{\mu\nu}\nabla_{\alpha}u^{\alpha}$
does not transform homogeneously. Based on Eq. (\ref{eq:transformation of u}),
we can get \citep{Baier:2007ix,Bhattacharyya:2007vs,Loganayagam:2008is}
\begin{eqnarray}
\nabla{}^{\left\langle \mu\right.}u^{\left.\nu\right\rangle } & = & \tau^{-3}\hat{\nabla}{}^{\left\langle \mu\right.}\hat{u}^{\left.\nu\right\rangle },\nonumber \\
\nabla_{\perp}^{\left[\mu\right.}u^{\left.\nu\right]} & = & \tau^{-3}\hat{\nabla}_{\perp}^{\left[\mu\right.}\hat{u}^{\left.\nu\right]},
\end{eqnarray}
which lead to a compact form for $\hat{\pi}^{\mu\nu}$ and $\hat{\phi}^{\mu\nu}$,
\begin{eqnarray}
\hat{\pi}^{\mu\nu}=\tau^{6}\pi^{\mu\nu} & = & -\hat{\eta}_{s}\hat{\nabla}{}^{\left\langle \mu\right.}\hat{u}^{\left.\nu\right\rangle }\nonumber \\
\hat{\phi}^{\mu\nu}=\tau^{6}\phi^{\mu\nu} & = & -\hat{\gamma}\left(\hat{\nabla}_{\perp}^{\left[\mu\right.}\hat{u}^{\left.\nu\right]}-2\hat{\Delta}^{\mu\alpha}\hat{\Delta}^{\nu\beta}\hat{\omega}_{\alpha\beta}\right),\label{eq:constitutive relations in ds3}
\end{eqnarray}
where 
\begin{equation}
\hat{\gamma}=\tau^{3}\gamma,\qquad\hat{\eta}_{s}=\tau^{3}\eta_{s}.\label{eq:gamma_ds3}
\end{equation}
Therefore, $\hat{\pi}^{\mu\nu}$ and $\hat{\phi}^{\mu\nu}$ have the
same structure as $\pi^{\mu\nu}$ and $\phi^{\mu\nu}$. Note that
we deliberately write $\hat{\eta}_{s}$ and $\hat{\gamma}$ as $(\hat{\eta}_{s}/\hat{s})\hat{s}$
and $(\hat{\gamma}/\hat{s})\hat{s}$, respectively. The $\hat{\eta}_{s}/\hat{s}$
and $\hat{\gamma}/\hat{s}$ are dimensionless scalars which do not
be modified when passing from Minkowski space-time $\mathbb{R}^{3,1}$
to $dS_{3}\times\mathbb{R}$ space-time. We follow the standard strategy
in Gubser flows and set 
\begin{equation}
\frac{\hat{\eta}_{s}}{\hat{s}}=\frac{\eta}{s}=\text{constant.,}\qquad\frac{\hat{\gamma}}{\hat{s}}=\frac{\gamma}{s}=\textrm{constant.}\label{eq:eta_s_01}
\end{equation}

Unfortunately, $\hat{q}^{\mu}$ becomes 
\begin{eqnarray}
\hat{q}^{\mu}=\tau^{5}q^{\mu} & = & -\hat{\lambda}\left(\frac{1}{\hat{T}}\hat{\nabla}_{\perp}^{\mu}\hat{T}-\hat{u}^{\alpha}\hat{\nabla}_{\alpha}\hat{u}^{\mu}-4\hat{\omega}^{\mu\nu}\hat{u}_{\nu}-2\tau^{-1}\hat{\Delta}^{\mu\alpha}\hat{\nabla}_{\alpha}\tau\right),\label{eq: q in ds3}
\end{eqnarray}
where $\hat{\lambda}=\tau^{3}\lambda$. The last term in the bracket
$-2\tau^{-1}\hat{\Delta}^{\mu\alpha}\hat{\nabla}_{\alpha}\tau$ is
generated by the Weyl rescaling. For simplicity, we need to set $\lambda=\hat{\lambda}=0$
in the current work, i.e. we set 
\begin{equation}
q^{\mu}=\hat{q}^{\mu}=0.
\end{equation}
In fact, we have also checked that the nonzero $\hat{q}^{\mu}$ breaks
the Gubser symmetry and will change the velocity $\hat{u}_{\mu}=\left(-1,0,0,0\right)$
due to the extra term $-2\tau^{-1}\hat{\Delta}^{\mu\alpha}\hat{\nabla}_{\alpha}\tau$.

In this section, we extend the conservation equations and constitutive
equations for the relativistic spin hydrodynamics from Minkowski space-time
$\mathbb{R}^{3,1}$ to the $dS_{3}\times\mathbb{R}$ space-time. We
find that neither the energy momentum conservation equation (\ref{eq:EM conservation in ds3})
nor the spin evolution equation (\ref{eq:spin evolution in ds3})
is covariant after Weyl rescaling. We also get the constitutive equations
in the $dS_{3}\times\mathbb{R}$ space-time shown in Eqs.(\ref{eq: T^=00003D00005Cmu=00003D00005Cnu-in ds3},
\ref{eq:constitutive relations in ds3}). We further set the bulk
viscosity $\zeta=0$ and $\hat{q}^{\mu}=0$ for simplicity.

\section{Analytic solutions in Gubser flow\label{sec:Analytic-solution-in}}

In this section, we derive the analytic solutions of dissipative spin
hydrodynamics in a Gubser flow in high temperature limit.

We adopt the strategy similar to our previous works \citep{Roy:2015kma,Pu:2016ayh,Pu:2016bxy,Pu:2016rdq,Siddique:2019gqh,Wang:2020qpx,Wang:2021ngp}.
First, we consider the thermodynamic relations and equations of state.
We express the energy density $\hat{e}$ and entropy density $\hat{s}$
in the $dS_{3}\times\mathbb{R}$ space-time as polynomial functions
of temperature$\hat{T}$ and spin chemical potential $\hat{\omega}^{\mu\nu}$
in Sec. \ref{subsec:Thermodynamic-relations}. Secondly, in Sec. \ref{subsec:Main-equations},
we concentrate on the fluid acceleration equations and search for
the special configurations for dissipative spin hydrodynamics, which
do not modify the fluid velocity (\ref{eq:fluid_velocity_01}) in
a Gubser flow. After that we succeed in finding self-consistent analytical
solutions for $\hat{e}$ and $\hat{S}^{\mu\nu}$. Finally, in Sec.
\ref{subsec:Analytic-solutions}, we convert the solutions obtained
in $dS_{3}\times\mathbb{R}$ space-time back to Minkowski space-time
$\mathbb{R}^{3,1}$ and compare them with the our solutions in a Bjorken
flow \citep{Wang:2021ngp}. We also discuss the results for the spin
hydrodynamics in the Belinfante form in Sec. \ref{sec:Results-for-Belinfante}.
Throughout this section, we use the Gubser coordinates $(\rho,\theta,\phi,\eta)$
in $dS_{3}\times\mathbb{R}$ space-time if not specified.

\subsection{Thermodynamic relations and equations of motion \label{subsec:Thermodynamic-relations}}

According to Eq. (\ref{eq: thermodynamic relations in R31}) and Eq.
(\ref{eq:transformation rules}), we rewrite the thermodynamic relations
in $dS_{3}\times\mathbb{R}$ space-time, 
\begin{eqnarray}
\hat{e}+\hat{p} & = & \hat{T}\hat{s}+\hat{\omega}_{\mu\nu}\hat{S}^{\mu\nu},\nonumber \\
d\hat{p} & = & \hat{s}d\hat{T}+\hat{S}^{\mu\nu}d\hat{\omega}_{\mu\nu}.\label{eq: thermodynamic relations in ds3}
\end{eqnarray}
Again, for simplicity, we set the number density and chemical potential
be zero. Following the standard Gubser flows \citep{Gubser:2010ze,Gubser:2010ui},
we can assume that the thermodynamic variables $\hat{e},\hat{p},\hat{T},\hat{s}$
and transport coefficients $\hat{\gamma},\hat{\eta}_{s}$ are only
functions of de Sitter time $\rho$. It suggests a natural assignment
that $\hat{\omega}_{\mu\nu}\hat{S}^{\mu\nu}$ depends on $\rho$ only.
We emphasize that due to the nontrivial metric $\hat{g}_{\mu\nu}=\textrm{diag}\{-1,\cosh^{2}\rho,\cosh^{2}\rho\sin^{2}\theta,1\}$
in the $dS_{3}\times\mathbb{R}$ space-time, $\hat{S}^{\mu\nu}$ and
$\hat{\omega}^{\mu\nu}$ may be the functions of both $\rho$ and
$\theta$.

To close the system, we need the equations of state besides Eq. (\ref{eq: thermodynamic relations in ds3}).
In \ref{subsec:Main-equations-in}, we assume that 
\begin{eqnarray}
\hat{e} & = & c_{s}^{-2}\hat{p}=3\hat{p},\label{eq: Eos-1}
\end{eqnarray}
which is a reasonable approximation in the ultra-relativistic or high
temperature limits. Here, $c_{s}^{2}$ is the speed of sound and usually
one can choose $c_{s}^{2}=1/3$ for simplicity. We emphasize that
EoS (\ref{eq: Eos-1}) does not imply the system is conformal invariant.
In fact, there is no the conformal symmetry in our system. More discussion
will be shown in the next subsection. On the other hand, inspired
by the relation between particle number density and chemical potential,
we assume another equation of state in high temperature limit \cite{Wang:2021ngp},
i.e., 
\begin{eqnarray}
\hat{S}^{\mu\nu} & = & a\hat{T}^{2}\hat{\omega}^{\mu\nu},\label{eq: Eos-2}
\end{eqnarray}
with dimensionless constant $a$. Eqs. (\ref{eq: Eos-1}, \ref{eq: Eos-2})
are regarded as two given conditions in the subsequent discussion.

For convenience, we further define a new auxiliary variable $\overline{\omega}^{2}$,
\begin{eqnarray}
\overline{\omega}^{2} & \equiv & \frac{\omega^{\mu\nu}\omega_{\mu\nu}}{T^{2}}=\frac{\hat{\omega}^{\mu\nu}\hat{\omega}_{\mu\nu}}{\hat{T}^{2}},\label{eq:omega_bar}
\end{eqnarray}
which is a dimensionless scalar and invariant under Weyl rescaling.

Utilizing these equations of state and transformation rule Eq. (\ref{eq:transformation rules}),
we can rewrite the thermodynamic relations Eq. (\ref{eq: thermodynamic relations in ds3})
as 
\begin{eqnarray}
\frac{4}{3}\hat{e} & = & \hat{T}\hat{s}+a\hat{T}^{4}\overline{\omega}^{2},\nonumber \\
\frac{1}{3}d\hat{e} & = & \left(\hat{s}+a\hat{T}^{3}\overline{\omega}^{2}\right)d\hat{T}+\frac{1}{2}a\hat{T}^{4}d\left(\overline{\omega}^{2}\right).\label{eq: New thermodynamic relations}
\end{eqnarray}
From Eq. (\ref{eq: New thermodynamic relations}), one can express
$\hat{e}=\hat{e}(\hat{T},\overline{\omega}^{2})$ and $\hat{s}=\hat{s}(\hat{T},\overline{\omega}^{2})$
as, 
\begin{eqnarray}
\hat{e} & = & \hat{T}^{4}\left(c_{0}+\frac{3}{2}a\overline{\omega}^{2}\right),\label{eq: expression for e}\\
\hat{s} & = & \hat{T}^{3}\left(\frac{4}{3}c_{0}+a\overline{\omega}^{2}\right),\label{eq: expression for s}
\end{eqnarray}
where 
\begin{equation}
c_{0}\equiv\frac{\hat{e}_{0}}{\hat{T}_{0}^{4}}-\frac{3}{2}a\overline{\omega}_{0}^{2}=\frac{3}{4}\frac{\hat{s}_{0}}{\hat{T}_{0}^{3}}-\frac{3}{4}a\overline{\omega}_{0}^{2},
\end{equation}
is a constant determined by initial values $\hat{e}_{0}=\hat{e}(\rho_{0})$,
$\hat{s}_{0}=\hat{s}(\rho_{0})$, $\hat{T}_{0}=\hat{T}(\rho_{0})$
and $\overline{\omega}_{0}^{2}=\overline{\omega}^{2}(\rho_{0})$.

\subsection{Simplify the differential equations \label{subsec:Main-equations}}

In this subsection, our task is to find special configuration to hold
the fluid velocity in a Gubser flow and simplify main differential
equations (\ref{eq:EM conservation in ds3}, \ref{eq:spin evolution in ds3}).

Contracting the projector $\hat{\Delta}_{\alpha\nu}=\hat{g}_{\alpha\nu}-\hat{u}_{\alpha}\hat{u}_{\nu}$
in the $dS_{3}\times\mathbb{R}$ space-time with both sides of Eq.
(\ref{eq:EM conservation in ds3}), yields the acceleration equation
for the fluid velocity, 
\begin{eqnarray}
\hat{\Delta}_{\alpha\nu}\left[\hat{\nabla}_{\mu}\hat{T}^{\mu\nu}-2\tau^{-1}\hat{T}^{[\mu\nu]}\hat{\nabla}_{\mu}\tau\right] & = & 0.
\end{eqnarray}
Plugging Eq. (\ref{eq: T^=00003D00005Cmu=00003D00005Cnu-in ds3})
into it, we get, 
\begin{eqnarray}
\hat{u}^{\mu}\hat{\nabla}_{\mu}\hat{u}_{\alpha} & = & -\frac{1}{\hat{e}+\hat{p}}\left[\hat{\Delta}_{\alpha}^{\mu}\hat{\nabla}_{\mu}\hat{p}+\hat{\Delta}_{\nu\alpha}\hat{\nabla}_{\mu}\hat{\pi}^{\mu\nu}+\hat{\Delta}_{\nu\alpha}\hat{\nabla}_{\mu}\hat{\phi}^{\mu\nu}\right.\left.-2\tau^{-1}\hat{g}_{\alpha\nu}\hat{\phi}^{\mu\nu}\hat{\nabla}_{\mu}\tau\right].\label{eq:acc_velocity_02}
\end{eqnarray}

To compute Eq. (\ref{eq:acc_velocity_02}), we find that only six
Christoffel symbols $\hat{\Gamma}_{\mu\nu}^{\lambda}$ in Gubser coordinates
$(\rho,\theta,\varphi,\eta)$ is nonzero, 
\begin{eqnarray}
\hat{\Gamma}_{\theta\theta}^{\rho}=\cosh\rho\sinh\rho,\; & \hat{\Gamma}_{\varphi\varphi}^{\rho}=\cosh\rho\sinh\rho\sin^{2}\theta,\; & \hat{\Gamma}_{\rho\theta}^{\theta}=\tanh\rho,\nonumber \\
\hat{\Gamma}_{\varphi\varphi}^{\theta}=-\sin\theta\cos\theta,\; & \hat{\Gamma}_{\rho\varphi}^{\varphi}=\tanh\rho,\; & \hat{\Gamma}_{\theta\varphi}^{\varphi}=\cot\theta.
\end{eqnarray}
Then, it is straightforward to get he nonzero components of $\hat{\pi}^{\mu\nu}$
and $\hat{\phi}^{\mu\nu}$ from Eq. (\ref{eq:constitutive relations in ds3}),
\begin{eqnarray}
\hat{\pi}^{\theta\theta} & = & -\frac{1}{3}\hat{\eta}_{s}\cosh^{-2}\rho\tanh\rho,\nonumber \\
\hat{\pi}^{\varphi\varphi} & = & -\frac{1}{3}\hat{\eta}_{s}\cosh^{-2}\rho\sin^{-2}\theta\tanh\rho,\nonumber \\
\hat{\pi}^{\eta\eta} & = & \frac{2}{3}\hat{\eta}_{s}\tanh\rho,\nonumber \\
\hat{\phi}^{ij} & = & 2\hat{\gamma}\hat{\omega}^{ij},\qquad(i,j=\theta,\varphi,\eta).\label{eq:pi_exp_01}
\end{eqnarray}
With the assumption that $\hat{e}$ and $\hat{p}$ depend on $\rho$
only, which leads to $\hat{\Delta}_{\alpha}^{\mu}\hat{\nabla}_{\mu}\hat{p}(\rho)=0$,
Eq. (\ref{eq:acc_velocity_02}) then becomes, 
\begin{eqnarray}
\hat{u}^{\mu}\hat{\nabla}_{\mu}\hat{u}_{\rho} & = & 0,\label{eq: acceleration-1}\\
\hat{u}^{\mu}\hat{\nabla}_{\mu}\hat{u}_{\theta} & = & 0,\\
\hat{u}^{\mu}\hat{\nabla}_{\mu}\hat{u}_{\varphi} & = & \frac{2\cosh^{2}\rho\sin^{2}\theta}{\hat{e}+\hat{p}}\left(\frac{\hat{\gamma}}{\hat{s}}\right)\hat{s}\left(-\partial_{\theta}\hat{\omega}^{\theta\varphi}-\cot\theta\hat{\omega}^{\theta\varphi}+2\hat{\omega}^{\theta\varphi}\tau^{-1}\partial_{\theta}\tau\right),\\
\hat{u}^{\mu}\hat{\nabla}_{\mu}\hat{u}_{\eta} & = & \frac{2}{\hat{e}+\hat{p}}\left(\frac{\hat{\gamma}}{\hat{s}}\right)\hat{s}\left(-\partial_{\theta}\hat{\omega}^{\theta\eta}-\cot\theta\hat{\omega}^{\theta\eta}+2\hat{\omega}^{\theta\eta}\tau^{-1}\partial_{\theta}\tau\right).\label{eq: acceleration-4}
\end{eqnarray}
Obviously, when $\hat{\omega}^{\theta\varphi},\hat{\omega}^{\theta\eta}=0$,
$\hat{u}^{\mu}\hat{\nabla}_{\mu}\hat{u}_{\nu}=0$, i.e. the Gubser
velocity (\ref{eq:fluid_velocity_01}) holds during the evolution
if $\hat{\omega}^{\theta\varphi},\hat{\omega}^{\theta\eta}$ always
vanish. Later, we will check that $\hat{\omega}^{\theta\varphi},\hat{\omega}^{\theta\eta}$
vanish under appropriate initial conditions and Gubser velocity (\ref{eq:fluid_velocity_01}).

Contacting $\hat{u}_{\nu}$ with both sides of Eq. (\ref{eq:EM conservation in ds3})
provides the conservation equation for energy, 
\begin{eqnarray}
\hat{u}_{\nu}\left[\hat{\nabla}_{\mu}\hat{T}^{\mu\nu}-2\tau^{-1}\hat{T}^{[\mu\nu]}\hat{\nabla}_{\mu}\tau\right] & = & 0.\label{eq:energy_ds3_01}
\end{eqnarray}
Using Eqs. (\ref{eq: T^=00003D00005Cmu=00003D00005Cnu-in ds3}, \ref{eq: Eos-1},
\ref{eq:pi_exp_01}), the evolution of energy density (\ref{eq:energy_ds3_01})
reads 
\begin{eqnarray}
\frac{d\hat{e}}{d\rho}+\frac{8}{3}\hat{e}\tanh\rho-\frac{2}{3}\left(\frac{\hat{\eta}_{s}}{\hat{s}}\right)\hat{s}\tanh^{2}\rho & = & 0.\label{eq: energy equation}
\end{eqnarray}
Eq. (\ref{eq: energy equation}) is the same as the one in ordinary
relativistic hydrodynamics without spin effect in a Gubser flow \citep{Gubser:2010ui,Gubser:2010ze}
(also see Refs. \citep{Yan:2015lfa,Martinez:2017ibh} for extensions).

Third, we compute the evolution of spin following Eq. (\ref{eq:spin evolution in ds3}).
After a long and tedious calculation, we eventually obtain six independent
equations for the evolution of spin from Eq. (\ref{eq:spin evolution in ds3}),
\begin{eqnarray}
\partial_{\rho}\hat{S}^{\rho\varphi}+3\tanh\rho\hat{S}^{\rho\varphi}+\hat{S}^{\theta\varphi}\tau^{-1}\partial_{\theta}\tau & = & 0,\label{eq:S_rho_phi}\\
\partial_{\rho}\hat{S}^{\rho\eta}+2\tanh\rho\hat{S}^{\rho\eta}+\hat{S}^{\theta\eta}\tau^{-1}\partial_{\theta}\tau & = & 0,\label{eq:S_rho_eta}\\
\partial_{\rho}\hat{S}^{\theta\varphi}+4\tanh\rho\hat{S}^{\theta\varphi}+\cosh^{-2}\rho\hat{S}^{\rho\varphi}\tau^{-1}\partial_{\theta}\tau+4\left(\frac{\hat{\gamma}}{\hat{s}}\right)\hat{s}\hat{\omega}^{\theta\varphi} & = & 0,\label{eq:S_theta_phi}\\
\partial_{\rho}\hat{S}^{\theta\eta}+3\tanh\rho\hat{S}^{\theta\eta}+\cosh^{-2}\rho\hat{S}^{\rho\eta}\tau^{-1}\partial_{\theta}\tau+4\left(\frac{\hat{\gamma}}{\hat{s}}\right)\hat{s}\hat{\omega}^{\theta\eta} & = & 0,\label{eq:S_theta_eta}
\end{eqnarray}
and, 
\begin{eqnarray}
\partial_{\rho}\hat{S}^{\varphi\eta}+3\tanh\rho\hat{S}^{\varphi\eta}+4\left(\frac{\hat{\gamma}}{\hat{s}}\right)\hat{s}\hat{\omega}^{\varphi\eta} & = & 0,\label{eq: spin evolution equation-1}\\
\partial_{\rho}\hat{S}^{\rho\theta}+3\tanh\rho\hat{S}^{\rho\theta} & = & 0.\label{eq: spin evolution equation-2}
\end{eqnarray}
As mentioned in Eq. (\ref{eq: acceleration-4}), the fixed Gubser
velocity requires that $\hat{\omega}^{\theta\varphi},\hat{\omega}^{\theta\eta}=0$.
This requirement leads that all $\hat{S}^{\theta\varphi},\hat{S}^{\theta\eta}$
should always be zero during the evolution through the EoS (\ref{eq: Eos-2}).
Unfortunately, $\hat{S}^{\theta\varphi},\hat{S}^{\theta\eta}$ are
coupled to $\hat{S}^{\rho\varphi},\hat{S}^{\rho\eta}$ through Eqs.(\ref{eq:S_rho_phi},\ref{eq:S_rho_eta},\ref{eq:S_theta_phi},\ref{eq:S_theta_eta}
). To satisfy the requirement from Eq. (\ref{eq: acceleration-4}),
we have to choose the trivial solutions of Eqs. (\ref{eq:S_rho_phi},\ref{eq:S_rho_eta},\ref{eq:S_theta_phi},\ref{eq:S_theta_eta}),
which is $\hat{S}^{\theta\varphi}(\rho,\theta)=\hat{S}^{\theta\eta}(\rho,\theta)=0$
and $\hat{S}^{\rho\varphi}(\rho,\theta)=\hat{S}^{\rho\eta}(\rho,\theta)=0$.

Remarkably, in the space-time $dS_{3}\times\mathbb{R}$, there are
extra terms proportional to $\hat{\nabla}_{\alpha}\tau$ in both energy-momentum
conservation equation (\ref{eq:EM conservation in ds3}) and the evolution
equations for spin (\ref{eq:spin evolution in ds3}). As mentioned
in the previous subsection, these terms come from the Weyl rescaling
and cannot be neglected in general. Fortunately, in the configuration
for the Gubser flow, all of these terms vanish in Eqs. (\ref{eq: energy equation},
\ref{eq: spin evolution equation-1}, \ref{eq: spin evolution equation-2}).
It is of great help for us to derive the analytic solutions in the
relativistic spin hydrodynamics in a Gubser flow. At last, we only
have three independent differential equations, i.e. conservation equation
for energy (\ref{eq: energy equation}) and evolution equations for
spin (\ref{eq: spin evolution equation-1}, \ref{eq: spin evolution equation-2}).

\subsection{Analytic solutions in $dS_{3}\times\mathbb{R}$ and $\mathbb{R}^{3,1}$
space-time \label{subsec:Analytic-solutions}}

In this subsection, we solve the differential equations (\ref{eq: energy equation},\ref{eq: spin evolution equation-1},\ref{eq: spin evolution equation-2})
for the spin hydrodynamics in a Gubser flow. We then transform our
solutions in $dS_{3}\times\mathbb{R}$ space-time to the Minkowski
space-time $\mathbb{R}^{3,1}$.

We consider the high temperature limit and the spin chemical potential
is much smaller than temperature in the relativistic heavy ion collisions,
i.e. $\omega^{\mu\nu}\ll T$, or 
\begin{equation}
\overline{\omega}^{2}=\frac{\omega^{\mu\nu}\omega_{\mu\nu}}{T^{2}}=\frac{\hat{\omega}^{\mu\nu}\hat{\omega}_{\mu\nu}}{\hat{T}^{2}}\ll1.\label{eq:power_01}
\end{equation}
We emphasize again that $\hat{\eta}_{s}/\hat{s}$ and $\hat{\gamma}/\hat{s}$
are small constants and we can assume $\hat{\eta}_{s}/\hat{s},\hat{\gamma}/\hat{s}\ll1$.
Therefore, we can consider the $\overline{\omega}^{2}$ and $\hat{\eta}_{s}/\hat{s},\hat{\gamma}/\hat{s}$
as small parameters and expand all the quantities in the power series
of $\overline{\omega}^{2}$ and $\hat{\eta}_{s}/\hat{s},\hat{\gamma}/\hat{s}$.

In leading order of $\overline{\omega}^{2}$, the Eq. (\ref{eq: energy equation})
becomes, 
\begin{eqnarray}
\frac{d\hat{T}}{d\rho}+\frac{2}{3}\hat{T}\tanh\rho-\frac{2}{9}\left(\frac{\hat{\eta}_{s}}{\hat{s}}\right)\tanh^{2}\rho+\mathcal{O}\left(\overline{\omega}^{2}\right) & = & 0,\\
\partial_{\rho}\hat{S}^{\varphi\eta}+3\tanh\rho\hat{S}^{\varphi\eta}+\frac{4}{a\hat{T}^{2}}\left(\frac{\hat{\gamma}}{\hat{s}}\right)\hat{s}\hat{S}^{\varphi\eta} & = & 0,\\
\partial_{\rho}\hat{S}^{\rho\theta}+3\tanh\rho\hat{S}^{\rho\theta} & = & 0,
\end{eqnarray}
whose solution is given by, 
\begin{eqnarray}
\hat{T} & = & \hat{T}_{0}\left(\frac{\cosh\rho_{0}}{\cosh\rho}\right)^{\frac{2}{3}}\left(1+\frac{\hat{\eta}_{s}}{\hat{s}}B(\rho)\right)+\mathcal{O}\left(\overline{\omega}^{2}\right),\label{eq: T in ds3}
\end{eqnarray}
with initial value $\hat{T}_{0}\equiv\hat{T}(\rho_{0})$. Here, the
auxiliary function $B(\rho)$ is 
\begin{eqnarray}
B(\rho) & \equiv & \frac{2}{27}\frac{1}{\hat{T}_{0}}\cosh^{-\frac{2}{3}}\rho_{0}\left[\sinh^{3}\rho F\left(\frac{7}{6},\frac{3}{2};\frac{5}{2};-\sinh^{2}\rho\right)\right.\nonumber \\
 &  & \qquad\left.-\sinh^{3}\rho_{0}F\left(\frac{7}{6},\frac{3}{2};\frac{5}{2};-\sinh^{2}\rho_{0}\right)\right],
\end{eqnarray}
where $F(a,b;c;z)$ is the hyper-geometric function. Substituting
Eq. (\ref{eq: T in ds3}) to Eq. (\ref{eq: expression for e}), we
obtain the expression for the energy density $\hat{e}$, 
\begin{eqnarray}
\hat{e} & = & \hat{e}_{0}\left(\frac{\cosh\rho_{0}}{\cosh\rho}\right)^{\frac{8}{3}}\left(1+4\frac{\hat{\eta}_{s}}{\hat{s}}B(\rho)\right)+\mathcal{O}\left(\overline{\omega}^{2},\left(\frac{\hat{\eta}_{s}}{\hat{s}}\right)^{2}\right).\label{eq: solution-e in ds3}
\end{eqnarray}

With the EoS (\ref{eq: Eos-2}), the solutions of evolution equations
for spin (\ref{eq: spin evolution equation-1}, \ref{eq: spin evolution equation-2})
are, 
\begin{eqnarray}
\hat{S}^{\rho\theta} & = & c_{1}\cosh^{-3}\rho,\label{eq: solution-S^/mu/nu in ds3-1}\\
\hat{S}^{\varphi\eta} & = & f(\theta)\cosh^{-3}\rho A(\rho)+\mathcal{O}\left(\overline{\omega}^{2}\right).\label{eq: solution-S^/mu/nu in ds3-2}
\end{eqnarray}
where 
\begin{eqnarray}
c_{1} & = & \hat{S}^{\rho\theta}(\rho_{0},\theta_{0})\cosh^{3}\rho_{0},\label{eq:def_c1}
\end{eqnarray}
is a constant determined by the initial condition and 
\begin{equation}
A(\rho)\equiv\exp\left[-\frac{4}{a}\int_{\rho_{0}}^{\rho}d\rho^{\prime}\left(\frac{\hat{\gamma}}{\hat{s}}\right)\frac{\hat{s}(\rho^{\prime})}{\hat{T}(\rho^{\prime})^{2}}\right].\label{eq:A_exp_rho}
\end{equation}
The $f(\theta)$ in Eq. (\ref{eq: solution-S^/mu/nu in ds3-2}) is
a function of $\theta$. As explained in Sec. \ref{subsec:Thermodynamic-relations},
although the Gubser flow requires the scalars $\hat{\omega}_{\mu\nu}\hat{S}^{\mu\nu}$
and both $\hat{\omega}_{\mu\nu}\hat{\omega}^{\mu\nu}$ and $\hat{S}_{\mu\nu}\hat{S}^{\mu\nu}$
depend on $\rho$ only, EoS (\ref{eq: Eos-2}) implies that $\hat{\omega}^{\mu\nu}$
or $\hat{S}^{\mu\nu}$ could also depend on $\theta$ due to the metric
$\hat{g}_{\mu\nu}=\mathrm{diag}\left(-1,\cosh^{2}\rho,\cosh^{2}\rho\sin^{2}\theta,1\right)$.
Using Eq. (\ref{eq: solution-S^/mu/nu in ds3-1}), we find that 
\begin{eqnarray}
\hat{S}^{\mu\nu}\hat{S}_{\mu\nu} & = & -2\cosh^{-4}\rho[c_{1}^{2}-2f^{2}(\theta)\sin^{2}\theta A(\rho)^{2}]+\mathcal{O}\left(\overline{\omega}^{2}\right),
\end{eqnarray}
Unless 
\begin{equation}
f(\theta)=c_{2}\sin^{-1}\theta,
\end{equation}
with a constant 
\begin{equation}
c_{2}=\hat{S}_{0}^{\varphi\eta}A^{-1}(\rho_{0})\cosh^{3}\rho_{0}\sin\theta_{0},
\end{equation}
determined by initial value of $\hat{S}_{0}^{\varphi\eta}=\hat{S}^{\varphi\eta}(\rho_{0},\theta_{0})$,
the $\hat{S}^{\mu\nu}\hat{S}_{\mu\nu}$ would not be independent on
$\theta$. Finally, the expression for the spin density $\hat{S}^{\varphi\eta}$
becomes 
\begin{eqnarray}
\hat{S}^{\varphi\eta} & = & c_{2}\cosh^{-3}\rho\sin^{-1}\theta A(\rho)+\mathcal{O}\left(\overline{\omega}^{2}\right).\label{eq: solution-S^=00003D00005Cmu=00003D00005Cnu-1 in ds3}
\end{eqnarray}

If the dimensionless quantity $\hat{\gamma}/\hat{s}$ can be regarded
as a small constant, i.e. $\hat{\gamma}/\hat{s}\ll1$, we can obtain
\begin{eqnarray}
A(\rho) & = & 1+\left(\frac{\hat{\gamma}}{\hat{s}}\right)\frac{6}{a}\frac{\hat{s}_{0}}{\hat{T}_{0}^{2}}\cosh^{\frac{2}{3}}\rho_{0}\left[\text{Sech}^{\frac{2}{3}}\rho_{0}F\left(\frac{1}{3},\frac{1}{2};\frac{4}{3};\text{Sech}^{2}\rho_{0}\right)\right.\nonumber \\
 &  & \qquad\left.-\text{Sech}^{\frac{2}{3}}\rho F\left(\frac{1}{3},\frac{1}{2};\frac{4}{3};\text{Sech}^{2}\rho\right)\right]\nonumber \\
 &  & +\mathcal{O}\left(\frac{\hat{\omega}^{\alpha\beta}\hat{\omega}_{\alpha\beta}}{\hat{T}^{2}},\left(\frac{\hat{\eta}_{s}}{\hat{s}}\right)^{2},\left(\frac{\hat{\gamma}}{\hat{s}}\right)^{2},\frac{\hat{\gamma}\hat{\eta}_{s}}{\hat{s}^{2}}\right).
\end{eqnarray}

Next, we transform the analytic solutions (\ref{eq: T in ds3},\ref{eq: solution-e in ds3},\ref{eq: solution-S^/mu/nu in ds3-1},\ref{eq: solution-S^=00003D00005Cmu=00003D00005Cnu-1 in ds3})
in the $dS_{3}\times\mathbb{R}$ space-time to the Minkowski space-time
$\mathbb{R}^{3,1}$. From Eq. (\ref{eq:energy_density_ds3}), the
energy density in Minkowski space-time $\mathbb{R}^{3,1}$ is 
\begin{eqnarray}
e & = & \frac{\hat{e}_{0}}{\tau_{0}^{4}}\left(\frac{\tau_{0}}{\tau}\right)^{\frac{4}{3}}\left[\frac{G(L,\tau_{0},x_{\perp0})}{G(L,\tau,x_{\perp})}\right]^{\frac{4}{3}}\times\left(1+4\frac{\eta_{s}}{s}B(\rho)\right)+\mathcal{O}\left(\overline{\omega}^{2},\left(\frac{\eta_{s}}{s}\right)^{2}\right).\label{eq: energy in R31}
\end{eqnarray}
where we introduce the 
\begin{equation}
G(L,\tau,x_{\perp})\equiv4L^{2}\tau^{2}+(L^{2}-\tau^{2}+x_{\perp}^{2})^{2},\label{eq:def_G}
\end{equation}
with an adjustable parameter $L$ defined in Eq. (\ref{eq:CartesianCoordinatesInR31}),
and the $\tau_{0}$ and $x_{\perp0}$ stands for the initial proper
time and the transverse position $x_{\perp}$. Here, we have used
the identity 
\begin{equation}
\hat{\eta}_{s}/\hat{s}=\eta_{s}/s,
\end{equation}
since $\hat{\eta}_{s}/\hat{s}$ is a scalar under the Weyl rescaling.
Using the same method, it is straightforward to get the expression
for temperature $T$ from Eq. (\ref{eq: T in ds3}). The $T$ as a
function of $\tau,x_{\perp}$ is similar to the $e(\tau,x_{\perp})$.

Next, we take the Weyl rescaling and the coordinate transformation
to the spin density. By using Eq. (\ref{eq: CT-2},\ref{eq:transformation rules}),
the nonzero spin density $S^{\mu\nu}$ in the $(\tau,x_{\perp},\varphi,\eta)$
coordinate system of Minkowski space-time $\mathbb{R}^{3,1}$ is given
by 
\begin{eqnarray}
S^{\tau x_{\perp}} & = & c_{1}\frac{4L^{2}}{\tau}G(L,\tau,x_{\perp})^{-1},\nonumber \\
S^{\varphi\eta} & = & c_{2}\frac{4L^{2}}{x_{\perp}\tau^{2}}G(L,\tau,x_{\perp})^{-1}A(\rho).\label{eq:S_tau_x_01}
\end{eqnarray}
We find that the exponential factor $A(\rho)$ in Eq. (\ref{eq:A_exp_rho})
is always less than $1$. It means that dissipative effects $\propto\hat{\gamma}$
accelerate decaying. Furthermore, we express the spin density in Cartesian
coordinates $(t,x,y,z)$ by coordinate transformation Eq. (\ref{eq: CT-1}),
i.e., 
\begin{eqnarray}
S^{0x} & = & \frac{4L^{2}}{\tau}C_{+}G(L,\tau,x_{\perp})^{-1},\label{eq: spin in R31-1}\\
S^{0y} & = & \frac{4L^{2}}{\tau}C_{-}G(L,\tau,x_{\perp})^{-1},\\
S^{xz} & = & \frac{4L^{2}}{\tau}D_{+}G(L,\tau,x_{\perp})^{-1},\\
S^{yz} & = & \frac{4L^{2}}{\tau}D_{-}G(L,\tau,x_{\perp})^{-1}.\label{eq: spin in R31-4}
\end{eqnarray}
where we introduce that 
\begin{eqnarray}
C_{\pm}(t,x,y,z) & = & c_{1}\cosh\eta\cos\varphi\pm c_{2}\sinh\eta\sin\varphi A(\rho),\\
D_{\pm}(t,x,y,z) & = & -c_{1}\sinh\eta\cos\varphi\pm c_{2}\cosh\eta\sin\varphi A(\rho),
\end{eqnarray}
and $\eta,\varphi,\rho$ are the functions of $(t,x,y,z)$ given by
Eqs. (\ref{eq: CT-1},\ref{eq: CT-2}). The other two components,
$S^{0z}$ and $S^{xy}$ vanish.

Let us comment on our results here. In large $L$ limit, i.e., $x_{\perp},\tau\ll L$,
we have $G(L,\tau,x_{\perp})\sim L^{4}$ and the energy density and
temperature become, 
\begin{eqnarray}
e\propto\tau^{-4/3}, & \qquad & T\propto\tau^{-1/3}.\label{eq:e_T_tau}
\end{eqnarray}
The spin density in Eqs. (\ref{eq: spin in R31-1}-\ref{eq: spin in R31-4})
reduces to, 
\begin{equation}
S^{0x},S^{0y},S^{xz},S^{yz}\propto L^{-2}\tau^{-1}.\label{eq:S_tau}
\end{equation}
From EoS (\ref{eq: Eos-2}), the spin chemical potential $\omega^{\mu\nu}$
decays like 
\begin{equation}
\omega^{0x},\omega^{0y},\omega^{xz},\omega^{yz}\propto L^{-2}\tau^{-1/3}.\label{eq:omega_tau}
\end{equation}
The decay behavior of $e,T,$ in large $L$ limit is the same as those
in the spin hydrodynamics in a Bjorken expansion \citep{Wang:2021ngp}.
Meanwhile, due to the dissipative effects, the spin density and spin
potential can decay more rapidly. We find that the exponential factor
$A(\rho)$ in Eq. (\ref{eq:A_exp_rho}) is always less than $1$.
It means that dissipative effects $\propto\hat{\gamma}$ accelerate
decaying.

Notably, in a Bjorken flow, we only get the nonzero solutions for
the spin component $S^{xy}$ \citep{Wang:2021ngp}. Here, there are
four non-vanishing spin density components found in the current work
due to the radial expansion in a Gubser flow.

Before end this section, let us discuss the terms in the modified
Cooper-Frye formula. As discussed in recent works for shear induced
polarization \citep{Liu:2021uhn,Liu:2021nyg,Fu:2021pok,Becattini:2021suc,Becattini:2021iol}
(also see Refs. \citep{Shi:2020qrx,Liu:2021nyg,Buzzegoli:2021wlg}
for other terms related to spin density), the thermal vortical $\mathcal{\text{\ensuremath{\Omega}}}^{\mu\nu}$
and thermal shear tensor $\xi^{\mu\nu}$
\begin{equation}
\mathcal{\text{\ensuremath{\Omega}}}^{\mu\nu}\equiv\frac{1}{2}\nabla^{\nu}(u^{\mu}/T)-\frac{1}{2}\nabla^{\mu}(u^{\nu}/T),\qquad\xi^{\mu\nu}\equiv\frac{1}{2}\nabla^{\mu}(u^{\nu}/T)+\frac{1}{2}\nabla^{\nu}(u^{\mu}/T),
\end{equation}
and spin potential $\omega^{\mu\nu}=S^{\mu\nu}/(aT^{2})$ appear in
modified Cooper-Frye formula. 

From Eqs. (\ref{eq:velocity in Minkowski},\ref{eq:transformation rules},\ref{eq: T in ds3}),
we find the nonzero components of $\Omega^{\mu\nu}$ are $\mathcal{\text{\ensuremath{\Omega}}}^{\tau x_{\perp}}$
and $\mathcal{\text{\ensuremath{\Omega}}}^{x_{\perp}\tau}$. Note
that, in $dS_{3}\times\mathbb{R}$ space-time, the nonzero thermal
vortical tensors $\hat{\mathcal{\text{\ensuremath{\Omega}}}}^{\tau x_{\perp}}$
and $\hat{\Omega}^{x_{\perp}\tau}$ come from the space-time derivatives
of the temperature and the extra terms proportional to $\hat{\nabla}_{\mu}\tau$
shown in Eq. (\ref{eq:transformation of u}) from Weyl rescaling.
In the $\mathbb{R}^{3,1}$ space-time, the $\nabla_{\mu}u_{\nu}$
with $u_{\mu}$ given by Eq. (\ref{eq:velocity in Minkowski}) is
obviously nonzero and can contribute to $\mathcal{\text{\ensuremath{\Omega}}}^{\tau x_{\perp}}$
and $\Omega^{x_{\perp}\tau}$ . 

We now analyze the evolution behavior of thermal vortical and shear
tensors.  

In large $L$ but small $\eta/s,\gamma/s$ limits, by using Eqs. (\ref{eq:velocity in Minkowski},\ref{eq: Eos-2},\ref{eq: T in ds3},\ref{eq:S_tau_x_01}),
we notice that 
\begin{equation}
\cosh\rho\propto L\tau^{-1},\quad u^{\tau}\propto\tau^{0},\quad u^{x_{\perp}}\propto L^{-2}\tau x_{\perp},
\end{equation}
and obtain the evolution behavior of nonzero components of $\Omega^{\mu\nu}$
and $\xi^{\mu\nu}$ listed in Tab \ref{tab:EvolutionBehavior}.
\begin{table}
\caption{Evolution behavior of nonzero components for thermal vortical tensor
$\Omega^{\mu\nu}$ and shear tensor $\xi^{\mu\nu}$ and spin potential
$\omega^{\mu\nu}$ in $\mathbb{R}^{3,1}$ space-time in large $L$
but small $\eta/s,\gamma/s$ limits. \label{tab:EvolutionBehavior} }

\noindent \centering{}%
\begin{tabular}{c|c|c|c|c|c|c|c}
\hline 
Nonzero components & $\mathcal{\text{\ensuremath{\Omega}}}^{\tau x_{\perp}}$ & $\mathcal{\text{\ensuremath{\xi}}}^{\tau\tau}$ & $\mathcal{\text{\ensuremath{\xi}}}^{\tau x_{\perp}}$ & $\mathcal{\text{\ensuremath{\xi}}}^{x_{\perp}x_{\perp}}$ & $\xi^{\varphi\varphi}$ & $\xi^{\eta\eta}$ & $\omega^{0x},\omega^{0y},\omega^{xz},\omega^{yz}$\tabularnewline
\hline 
\hline 
Evolution behavior &  &  &  &  &  &  & \tabularnewline
in large $L$ and & $L^{-2}\tau^{1/3}$ & $L^{0}\tau^{-2/3}$ & $L^{-2}\tau^{1/3}$ & $L^{-2}\tau^{4/3}$ & $L^{-2}\tau^{4/3}$ & $L^{0}\tau^{-8/3}$ & $L^{-2}\tau^{-1/3}$\tabularnewline
small $\eta/s,\gamma/s$ limits &  &  &  &  &  &  & \tabularnewline
\hline 
\end{tabular}
\end{table}
 We find that the only nonzero component of thermal vortical tensor,
$\mathcal{\text{\ensuremath{\Omega}}}^{\tau x_{\perp}}$, is much
smaller than the maximum component of thermal shear tensor, $\xi^{\tau\tau}$,
but it has the same order of magnitude as the spin potential in Eq.
(\ref{eq:omega_tau}). 

From Eq. (\ref{eq:constitutive relation}), in the global equilibrium,
one of the most important conclusion for spin hydrodynamics is that
the $\nabla_{\perp}^{\left[\mu\right.}u^{\left.\nu\right]}-2\Delta^{\mu\alpha}\Delta^{\nu\beta}\omega_{\alpha\beta}=T[\Delta^{\mu\alpha}\Delta^{\nu\beta}\Omega_{\alpha\beta}-T^{-1}\Delta^{\mu\alpha}\Delta^{\nu\beta}\omega_{\alpha\beta}]$
should be zero \cite{Becattini_plb2019_bfs,Hattori:2019lfp,Fukushima:2020ucl}.
Here, we compare the evolution behavior of the following quantities,
\begin{align}
\Delta^{\mu\alpha}\Delta^{\nu\beta}\Omega_{\alpha\beta} & \propto L^{-2}\tau^{1/3},\nonumber \\
\Delta^{\mu\alpha}\Delta^{\nu\beta}\xi_{\alpha\beta} & \propto L^{0}\tau^{-2/3},\nonumber \\
T^{-1}\Delta^{\mu\alpha}\Delta^{\nu\beta}\omega_{\alpha\beta} & \propto L^{-2}\tau^{0},
\end{align}
We conclude that in large $L$ but small $\eta/s,\gamma/s$ limits
the thermal shear tensor is more important than than spin potential
and thermal vortical tensor. 

\begin{figure}
\begin{centering}
\includegraphics[scale=0.26]{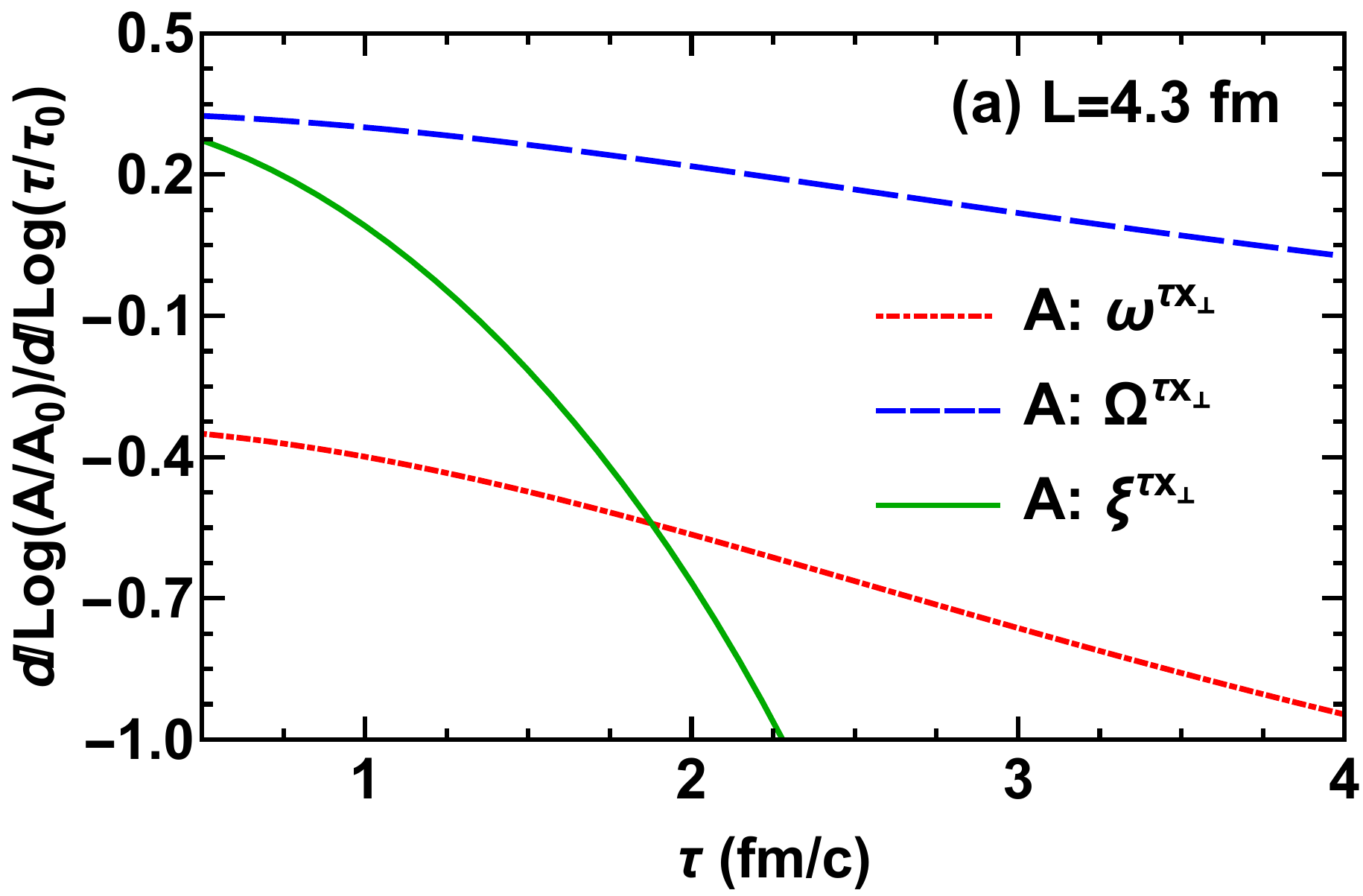} \includegraphics[scale=0.26]{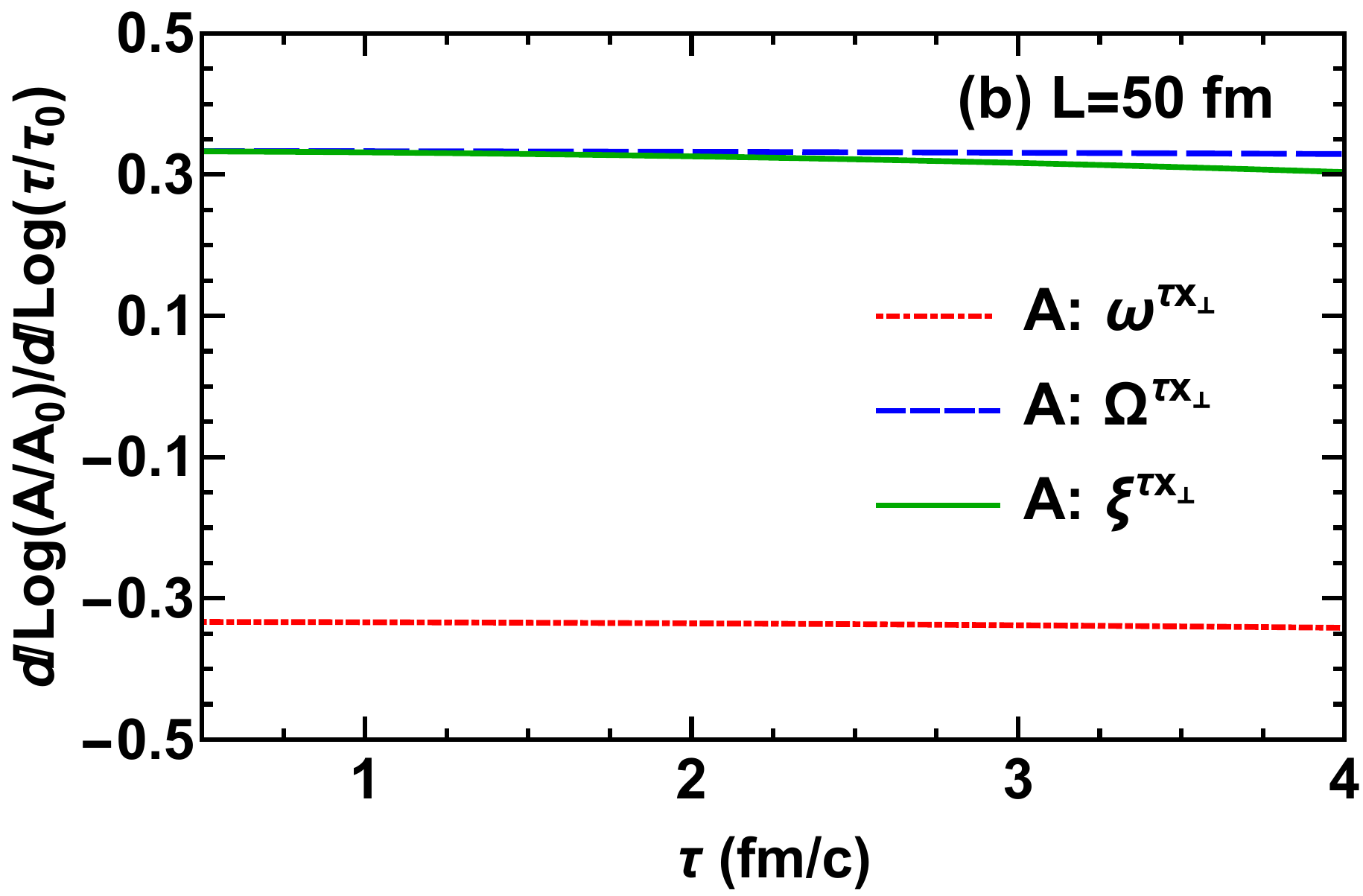}
\par\end{centering}
\caption{We plot $d[\log(A/A_{0})]/d[\log(\tau/\tau_{0})]$ with $A=\Omega^{\tau x_{\perp}},\xi^{\tau x_{\perp}},\omega^{\tau x_{\perp}}$
as functions of proper time $\tau$ at $x_{\perp}=0.5\textrm{fm}$
for (a) $L=4.3\textrm{fm}$ and (b) $L=50\textrm{fm}$. The parameters
are chosen as $\hat{e}_{0}=880$ at $\rho_{0}=0$ with $\eta_{s}=0$,
$c_{0}=11$ and $\hat{T}_{0}=(\hat{e}_{0}/c_{0})^{1/4}$. \citep{Gubser:2010ze,Gubser:2010ui}.The red dotted, blue dashed and green solid lines stand for the cases of  $A=\Omega^{\tau x_{\perp}},\xi^{\tau x_{\perp}},\omega^{\tau x_{\perp}}$, respectively. 
\label{fig:Decaying}}
\end{figure}

We also discuss the evolution behavior at finite $L$ case. We follow
the in-viscid case of Gubser flow in Ref. \citep{Gubser:2010ze,Gubser:2010ui}
and take $\hat{e}_{0}=880$ at $\rho_{0}=0$ with $\eta_{s}=0$, $c_{0}=11$
and $\hat{T}_{0}=(\hat{e}_{0}/c_{0})^{1/4}$ and $L=4.3\textrm{fm}$.
To see the power law behavior, we plot $d[\log(A/A_{0})]/d[\log(\tau/\tau_{0})]$
as a function of $\tau$. Here, we choose $A=\Omega^{\tau x_{\perp}},\xi^{\tau x_{\perp}},\omega^{\tau x_{\perp}}$
and $A_{0}$ stands for the value of these quantities at initial proper
time $\tau_{0}$. In Fig. \ref{fig:Decaying}, we choose $x_{\perp}=0.5\textrm{fm}$
for simplicity. We have checked numerically that power law behavior
of these quantities for other fixed $x_{\perp}$ ($\lesssim4\textrm{fm}$)
almost the same. The maximum proper time is chosen as $4\textrm{fm}/c$.
After $4\textrm{fm}/c$, the temperature will be less than the typical
freeze out one \citep{Gubser:2010ze,Gubser:2010ui}. 

In Fig. \ref{fig:Decaying}(a), we find that within $4\textrm{fm}/c$
the $\Omega^{\tau x_{\perp}}$ and $\omega^{\tau x_{\perp}}$ always
increases or decreases, respectively. While, the thermal shear tensor
$\xi^{\tau x_{\perp}}$ increases as $\propto\tau^{1/3}$ at early
time but decrease rapidly as $\propto\tau^{-1}$ after $2.3\textrm{fm}/c$.
Surprisingly, the thermal shear tensor decays much faster than the
spin potential in this model. As a comparison, when $L=50\textrm{fm}$
in Fig. \ref{fig:Decaying}(b), we observe a consistent results as
expected in Tab \ref{tab:EvolutionBehavior}. 

We conclude that in finite $L$ case, the evolution behavior of the
quantities mentioned above depends on the parameters $L$.  Therefore,
we can not naively drop any one of them in the modified Cooper-Frye
formula. To clarify it, further studies based on spin hydrodynamics
are needed in the future.

\subsection{Results for Belinfante form \label{sec:Results-for-Belinfante}}

Different with the canonical form of energy-momentum and angular momentum
tensor, one can define the energy-momentum tensor in Belinfante form
$\mathcal{T}^{\mu\nu}$ through the pseudogauge transformation, 
\begin{eqnarray}
\mathcal{T}^{\mu\nu} & = & T^{\mu\nu}+\partial_{\lambda}K^{\lambda\mu\nu},
\end{eqnarray}
where 
\begin{equation}
K^{\lambda\mu\nu}=\frac{1}{2}\left(\Sigma^{\lambda\mu\nu}-\Sigma^{\mu\lambda\nu}+\Sigma^{\nu\mu\lambda}\right),\label{eq:pesudo_gauge_01}
\end{equation}
and $\Sigma^{\lambda\mu\nu}$ is given by Eq. (\ref{eq:sigma_01}).
The Belinfante total angular momentum reads, 
\begin{equation}
\mathcal{J}^{\alpha\mu\nu}=J^{\alpha\mu\nu}+\partial_{\rho}(x^{\mu}K^{\rho\alpha\nu}-x^{\nu}K^{\rho\alpha\mu})=x^{\mu}\mathcal{T}^{\alpha\nu}-x^{\nu}\mathcal{T}^{\alpha\mu}.
\end{equation}
It is obviously that both $\mathcal{T}^{\mu\nu}$ and $\mathcal{J}^{\alpha\mu\nu}$
are conserved.

After a short calculation, up to $\mathcal{O}(\partial^{1})$, one
get \citep{Fukushima:2020ucl}, 
\begin{eqnarray}
\mathcal{T}^{\mu\nu} & = & eu^{\mu}u^{\nu}+p\Delta^{\mu\nu}+\frac{1}{2}\nabla_{\lambda}(u^{\mu}S^{\nu\lambda}+u^{\nu}S^{\mu\lambda})\nonumber \\
 & = & (e+\delta e_{s})u^{\mu}u^{\nu}+(p+\delta\Pi_{s})\Delta^{\mu\nu}+2\delta h_{s}^{(\mu}u^{\nu)}+(\pi^{\mu\nu}+\delta\pi_{s}^{\mu\nu}),
\end{eqnarray}
where 
\begin{eqnarray}
\delta e_{s} & = & -u_{\mu}\nabla_{\lambda}S^{\mu\lambda},\nonumber \\
\delta h_{s}^{\mu} & = & \frac{1}{2}(\Delta_{\beta}^{\mu}\nabla_{\lambda}S^{\beta\lambda}-u_{\beta}S^{\beta\lambda}\nabla_{\lambda}u^{\mu}),\nonumber \\
\delta\pi_{s}^{\mu\nu} & = & \nabla_{\lambda}(u^{\left\langle \mu\right.}S^{\left.\nu\right\rangle \lambda}),\nonumber \\
\delta\Pi_{s} & = & \frac{1}{3}\Delta_{\rho\sigma}\nabla_{\lambda}(u^{\rho}S^{\sigma\lambda}),\label{eq:delta_spin_01}
\end{eqnarray}
are the spin corrections to the energy density, heat flow, shear viscous
tensor and bulk viscous pressure, respectively. 

In our previous work \citep{Wang:2021ngp}, all of these spin corrections
vanish in a Bjorken flow. Inserting our results (\ref{eq:S_tau_x_01})
with Gubser velocity (\ref{eq:velocity in Minkowski}) into Eq. (\ref{eq:delta_spin_01})
yields, the spin corrections to the energy density, 
\begin{eqnarray}
\delta e_{s} & = & 4c_{1}\tau^{-1}x_{\perp}^{-1}L^{2}(L^{2}+\tau^{2}-3x_{\perp}^{2})G(L,\tau,x_{\perp})^{-3/2},\label{eq:de_01}
\end{eqnarray}
where $G$ and constant $c_{1}$ are given by Eqs. (\ref{eq:def_G})
and (\ref{eq:def_c1}), respectively. We get the other non-vanishing
spin corrections in the $(\tau,x_{\perp},\phi,\eta)$ coordinates,
\begin{eqnarray}
\delta h_{s}^{\tau} & = & 16c_{1}L^{2}\tau x_{\perp}G(L,\tau,x_{\perp})^{-2},\nonumber \\
\delta h_{s}^{x_{\perp}} & = & 8c_{1}L^{2}\left(L^{2}+x_{\perp}^{2}+\tau^{2}\right)G(L,\tau,x_{\perp})^{-2},\nonumber \\
\delta\pi_{s}^{\tau\tau} & = & -\frac{64}{3}c_{1}L^{2}\tau x_{\perp}^{3}G(L,\tau,x_{\perp})^{-5/2},\nonumber \\
\delta\pi_{s}^{x_{\perp}x_{\perp}} & = & -\frac{16}{3}c_{1}L^{2}\tau^{-1}x_{\perp}\left(L^{2}+x_{\perp}^{2}+\tau^{2}\right)^{2}G(L,\tau,x_{\perp})^{-5/2},\nonumber \\
\delta\pi_{s}^{\varphi\varphi} & = & \frac{8}{3}c_{1}L^{2}\tau^{-1}x_{\perp}^{-1}G(L,\tau,x_{\perp})^{-3/2},\nonumber \\
\delta\pi_{s}^{\eta\eta} & = & \frac{8}{3}c_{1}L^{2}\tau^{-3}x_{\perp}G(L,\tau,x_{\perp})^{-3/2},\nonumber \\
\delta\pi_{s}^{\tau x_{\perp}} & = & \delta\pi_{s}^{x_{\perp}\tau}=-\frac{32}{3}c_{1}L^{2}x_{\perp}^{2}\left(L^{2}+x_{\perp}^{2}+\tau^{2}\right)G(L,\tau,x_{\perp})^{-5/2},\nonumber \\
\delta\pi_{s}^{\varphi\eta} & = & \delta\pi_{s}^{\eta\varphi}=\frac{1}{2}S^{\varphi\eta}\tau^{-1}\left(L^{2}+x_{\perp}^{2}-\tau^{2}\right)G(L,\tau,x_{\perp})^{-1/2},\nonumber \\
\delta\Pi_{s} & = & -\frac{8}{3}c_{1}L^{2}\tau^{-1}x_{\perp}G(L,\tau,x_{\perp})^{-3/2}.\label{eq:dissipative_spin_correction}
\end{eqnarray}

One may wonder why the spin correction to the bulk pressure $\delta\Pi_{s}$
in Eq. (\ref{eq:dissipative_spin_correction}) is nonzero and may
break the conformal invariance. Again, we comment that EoS (\ref{eq: Eos-1})
is the leading order one in the ultra high temperature limits and
is not related to the conformal invariance directly. Therefore, there
is no inconsistency between the finite $\delta\Pi_{s}$ and EoS (\ref{eq: Eos-1}).
The $\delta\Pi_{s}$ comes from the pseudogauge transformation and
belong to the Belinfante energy momentum tensor $\mathcal{T}^{\mu\nu}$.
In the canonical form, if the initial bulk pressure is zero, bulk
pressure is always vanishing in the presence of spin potential. Since
the spin hydrodynamics in Belinfante form can be quite different with
those in canonical form (one can find examples in Ref.\cite{Fukushima:2020ucl}),
it is not surprising that we have different bulk pressure in two forms.

Now, we turn to estimate how large these spin corrections will be.
In the large $L$ limit, we find energy density $e\sim\tau^{-4/3}$
and its spin corrections $\delta e_{s}\sim1/(\tau x_{\perp}L^{2})$.
It gives that $\lim_{L\rightarrow\infty}\delta e_{s}/e\propto\tau^{1/3}/L^{2}\rightarrow0$
at late proper time, i.e. the spin correction to the energy density
$\delta e_{s}$ is a small correction to the energy density $e$. 

For a realistic model for Gubser flow, we choose $\hat{e}_{0}=(5.55)^{4}$
at $\rho_{0}=0$ with $\eta_{s}/s=0.268$, $c_{0}=11$,$\hat{T}_{0}=(\hat{e}_{0}/c_{0})^{1/4}$
and the parameter of characteristic length $L$ as $4.3\textrm{fm}$
\citep{Gubser:2010ze,Gubser:2010ui}. Note that, the due to the differences
of notations, $\eta_{s}/s$ in this work is twice as much as that
in Ref. \citep{Gubser:2010ze,Gubser:2010ui}. Since the proper time
is larger than $4.0\textrm{fm}/c$, the temperature is less than the
typical freeze out temperature $T\leq150\textrm{MeV}$. We choose
the range of proper time as $0.5-4.0\textrm{fm}/c$ similar to the
standard Gubser flow \citep{Gubser:2010ze,Gubser:2010ui}. Although
total energy correction $\int dx_{\perp}x_{\perp}\delta e_{s}$ is
finite, $\delta e_{s}\propto x_{\perp}^{-1}$ may be divergent as
$x_{\perp}\rightarrow0$. To avoid the divergent behavior of $\delta e$,
we have imposed the constraint $x_{\perp}>0.5\textrm{fm}$ in Eq.
(\ref{eq:deltae_e}). The parameter $c_{1}$ defined in Eq. (\ref{eq:def_c1})
should not be too large due to our power counting scheme in Eq. (\ref{eq:power_01})
in Sec. \ref{subsec:Analytic-solutions}. Here, we choose $|c_{1}|\leq2$
as a reasonable test. Using these parameters and Eqs. (\ref{eq:velocity in Minkowski},\ref{eq:constitutive relation},\ref{eq: Eos-1},\ref{eq: energy in R31},\ref{eq:de_01},\ref{eq:dissipative_spin_correction}),
we obtain
\begin{align}
\big|\delta e_{s}/e\big| & <0.1,\text{ at }x_{\perp}\in[0.5,4.0]\textrm{fm},\tau\in[0.5,4.0]\textrm{fm}/c,\label{eq:deltae_e}\\
\big|\delta\pi_{s}^{\mu\nu}/\pi^{\mu\nu}\big|,\big|\delta\Pi_{s}/p\big| & <0.1,\text{ at }x_{\perp}\in[0.0,4.0]\textrm{fm},\tau\in[0.5,4.0]\textrm{fm}/c.
\end{align}

We comment that Eqs.(\ref{eq:de_01}) and (\ref{eq:dissipative_spin_correction})
are the evidence to show that spin corrections in the Belinfante form
of spin hydrodynamic exist. These spin corrections are expected in
Ref.\cite{Fukushima:2020ucl}.

\section{Conclusion \label{sec:Conclusion}}

In this work we have obtained the analytical solutions for the dissipative
spin hydrodynamics with radial expansion in a Gubser flow.

After a short review on the standard Gubser flow, we briefly discuss
the relativistic dissipative spin hydrodynamics in the Minkowski space-time
$\mathbb{R}^{3,1}$ and extend the main equations to the $dS_{3}\times\mathbb{R}$
space-time under Weyl rescaling. Unfortunately, we find that there
are extra contributions $\propto\hat{\nabla}_{\mu}\tau$ from Weyl
rescaling to both the energy momentum conservation Eq. (\ref{eq:EM conservation in ds3})
and angular momentum conservation Eq.(\ref{eq:spin evolution in ds3}).
We emphasize that the energy-momentum conservation equations is no
longer conformal invariant in the current work due to its anti-symmetric
components. For simplicity, we drop the bulk viscous pressure and
$\hat{q}^{\mu}$. We further assume the transport coefficients $\hat{\eta}_{s}/\hat{s}$
and $\hat{\gamma}/\hat{s}$ are small constants similar to the ordinary
Gubser flow.

We then discuss the thermodynamic relations (\ref{eq: thermodynamic relations in ds3})
and the equations of state (\ref{eq: Eos-1},\ref{eq: Eos-2}) in
the $dS_{3}\times\mathbb{R}$ space-time. For convenience, we introduce
a dimensionless scalar $\overline{\omega}^{2}$ in Eq. (\ref{eq:omega_bar}).
Next, we derive the special configuration for the fluid, in which
the fluid velocity in a Gubser flow holds. Fortunately, the extra
terms from Weyl rescaling $\sim\hat{\nabla}_{\mu}\tau$ in energy
momentum and angular momentum conservation equations vanish in this
configuration. In the power series expansion of small $\overline{\omega}^{2},\hat{\eta}_{s}/\hat{s},\hat{\gamma}/\hat{s}$,
we have derived the analytic solutions for the dissipative spin hydrodynamics
in the $dS_{3}\times\mathbb{R}$ space-time. The evolution of energy
density and spin density $\hat{S}^{\rho\theta},\hat{S}^{\varphi\eta}$
are shown in Eqs. (\ref{eq: solution-e in ds3},\ref{eq: solution-S^/mu/nu in ds3-1},\ref{eq: solution-S^/mu/nu in ds3-2})
in $dS_{3}\times\mathbb{R}$ space-time. We also transform these physical
quantities back to the Minkowski space-time $\mathbb{R}^{3,1}$.

Our main results for the energy density $e$ and spin density $S^{0x},S^{0y},S^{xz},S^{yz}$
in Minkowski space-time $\mathbb{R}^{3,1}$ are given by Eqs. (\ref{eq: energy in R31},\ref{eq: spin in R31-1}-\ref{eq: spin in R31-4}).
There are two remarkable difference between the solutions found here
and in a Bjorken flow \citep{Wang:2021ngp}. The first thing is that
the solutions in a Gubser flow provide the additional information
for transverse expansion of the systems, which is missing in a Bjorken
flow \citep{Wang:2021ngp}. Meanwhile, now we have derived four nonzero
components $S^{0x},S^{0y},S^{xz},S^{yz}$ in spin density tensor in
current work, while we only have one nonzero component $S^{xy}$ in
Bjorken flow \citep{Wang:2021ngp}. It indicates that our current
findings are not a simple extension of Bjorken flow.

In large $L$ and small $\eta/s,\gamma/s$ limits we find that $e\propto\tau^{-4/3},T\propto\tau^{-1/3},S^{\mu\nu}\propto L^{-2}\tau^{-1},\omega^{\mu\nu}\propto L^{-2}\tau^{-1/3}$,
which are similar to the behavior in Bjorken expansion \citep{Wang:2021ngp}.
Moreover, $S^{\mu\nu}$ decay much faster in the cases of a finite
$L$ and with nonzero dissipative effects than in large $L$ and prefect
fluid limit.

In our model, we find that in large $L$ and small $\eta/s,\gamma/s$
limits the thermal shear tensor $\xi^{\mu\nu}$ may be more important
than spin potential $\omega^{\mu\nu}$ and thermal vortical tensor
$\mathcal{\text{\ensuremath{\Omega}}}^{\mu\nu}$, but in finite $L$
case their evolution behavior depends on the parameters $L$ strongly.
We can not naively drop any one of them in the modified Cooper-Frye
formula. To clarify it, we need further studies based on spin hydrodynamics
in the future.

At the end, after the pseudogauge transformation, we discuss the results
for spin hydrodynamics in the Belinfante form. We observe that the
spin corrections to the energy density and other dissipative terms
do not vanish in a Gubser flow, which is quite different with the
results in a Bjorken flow.

Our analytic solutions also provide the test beds for the future numerical
simulations of relativistic dissipative spin hydrodynamics . 
\begin{acknowledgments}
This work is partly supported by National Natural Science Foundation
of China (NSFC) under Grants No. 12075235 and 12135011. 
\end{acknowledgments}

 \bibliographystyle{h-physrev}
\bibliography{spinhydro}

\end{document}